\documentclass[journal,12pt,onecolumn,draftclsnofoot]{IEEEtran}
\usepackage{algpseudocode}
\usepackage{pict2e}
\usepackage{cite}
\usepackage[table]{xcolor}
\usepackage{supertabular}
\usepackage{booktabs}
\usepackage{tikz}

\usetikzlibrary{calc, shadings, shadows, shapes.arrows}
\usetikzlibrary{decorations.pathreplacing}
\usetikzlibrary{shapes}
\usepackage{amsmath}
\usepackage{nccmath}
\usepackage{amssymb}
\usepackage{amsthm}
\usepackage{bbm}
\usepackage[algoruled, linesnumbered]{algorithm2e}
\usepackage{algpseudocode}
\usepackage{graphicx}
\usepackage{caption}
\usepackage{subcaption}
\usepackage{verbatim}
\usepackage{longtable}
\usepackage{array}
\usepackage{arydshln}
\usepackage{lscape}
\usepackage{multirow}
\usepackage{makecell}
\usepackage{slashbox}	
\usepackage{enumitem}

\DeclareMathOperator*{\argmax}{arg\,max}

\setlist[itemize]{leftmargin=*}
\graphicspath{{Figs/}}
\theoremstyle{plain} 
\theoremstyle{plain} 
\theoremstyle{definition} 
\theoremstyle{definition} 
\theoremstyle{definition} 

\makeatletter
\def\bbordermatrix#1{\begingroup \m@th
	\@tempdima 4.75\p@
	\setbox\z@\vbox{%
		\def\cr{\crcr\noalign{\kern2\p@\global\let\cr\endline}}%
		\ialign{$##$\hfil\kern2\p@\kern\@tempdima&\thinspace\hfil$##$\hfil
			&&\quad\hfil$##$\hfil\crcr
			\omit\strut\hfil\crcr\noalign{\kern-\baselineskip}%
			#1\crcr\omit\strut\cr}}%
	\setbox\tw@\vbox{\unvcopy\z@\global\setbox\@ne\lastbox}%
	\setbox\tw@\hbox{\unhbox\@ne\unskip\global\setbox\@ne\lastbox}%
	\setbox\tw@\hbox{$\kern\wd\@ne\kern-\@tempdima\left[\kern-\wd\@ne
		\global\setbox\@ne\vbox{\box\@ne\kern2\p@}%
		\vcenter{\kern-\ht\@ne\unvbox\z@\kern-\baselineskip}\,\right]$}%
	\null\;\vbox{\kern\ht\@ne\box\tw@}\endgroup}
\makeatother

\title{A Dynamic Reliability-Aware Service Placement for Network Function Virtualization (NFV)}

\author{
	Mohammad Karimzadeh Farshbafan, Vahid Shah-Mansouri, and
	\thanks{
		Mohammad Karimzadeh Farshbafan and Vahid Shah-Mansouri are with School of Electrical and Computer Engineering, College of Engineering, University of Tehran, Tehran, Iran (e-mail: \{m.karimzadeh68,vmansouri\}@ut.ac.ir).
	}%
	Dusit Niyato,
	\thanks{
		Dusit Niyato is with School of Computer Science and Engineering, Nanyang Technological University, Singapore (e-mail: dniyato@ntu.edu.sg).
	}%
	\vspace{-1mm}

}

\allowdisplaybreaks[1]

\begin{document}

	\maketitle
	
	\begin{abstract}
     Network softwarization is one of the major paradigm shifts in the next generation of networks. It enables programmable and flexible management and deployment of the network. Network function virtualization (NFV) is referred to the deployment of software functions running on commodity servers instead of traditional hardware-based middle-boxes. It is an example of network softwarization. In NFV, a service is defined as a chain of software functions named service chain function (SFC). The process of allocating the resources of servers to the services, called service placement, is the most challenging mission in NFV. Dynamic nature of the service arrivals and departures as well as meeting the service level agreement make the service placement problem even more challenging. In this paper, we propose a model for dynamic reliability-aware service placement based on the simultaneous allocation of the main and backup servers. Then, we formulate the dynamic reliability-aware service placement as an infinite horizon Markov decision process (MDP), which aims to minimize the placement cost and maximize the number of admitted services. In the proposed MDP, the number of active services in the network is considered to be the state of the system, and the state of the idle resources is estimated based on it. Also, the number of possible admitted services is considered as the action of the presented MDP. To evaluate each possible action in the proposed MDP, we use a sub-optimal method based on the Viterbi algorithm named Viterbi-based Reliable Static Service Placement (VRSSP) algorithm. We determine the optimal policy based on value iteration method using an algorithm named VRSSP-based Value Iteration (VVI) algorithm. Eventually, through the extensive simulations, the superiority of the proposed model for dynamic reliability-aware service placement compared to the static solutions is inferred.
		
	\end{abstract}
	\vspace{-6mm}
\begin{IEEEkeywords}
\vspace{-3mm}
	Network function virtualization (NFV), dynamic reliability-aware service placement, Markov decision process (MDP), Viterbi algorithm, value iteration algorithm.
\end{IEEEkeywords}
\vspace{-6mm}
\section{Introduction}
\label{Section: Introduction}
Network softwarization technologies are envisioned to be a major contribution to 5G networks \cite{afolabi2018network}. In this way, network function virtualization (NFV) avoids the necessity of dedicated middleboxes and facilitates the agile service provisioning \cite{mijumbi2016network}. In the NFV paradigm, the hardware middleboxes are replaced by software-based virtual network functions (VNFs) which run on commodity servers. Examples of these VNFs are deep packet inspection (DPI), firewall, and cellular packet core functions \cite{laghrissi2018survey}. In NFV, a service is created by instantiation of multiple connected VNFs, called service function chain (SFC). The task of assigning SFCs and their VNFs to appropriate servers is referred to as service placement \cite{herrera2016resource}. The main components of an NFV-based network are
\begin{itemize}
\item \textbf{Infrastructure network provider} (InP) is the owner of the network function virtualization infrastructure (NFVI) which includes commodity servers for performing the VNFs and the links between the servers for routing the traffic of back-to-back VNFs in the SFCs.
\item \textbf{Services} are requested by the users of the network. Each service has a specific service level agreements (SLA) (e.g., the reliability of service or end-to-end delay). The SLA of each service is determined according to the type of the service.
\item \textbf{Network Operator} (NO) is responsible for providing the services of the users. For this purpose, NO first composes an appropriate SFC for each service and then use the infrastructure of the InP for performing service placement \cite{herrera2016resource}.
\end{itemize}

The most important challenge of a successful NFV deployment is managing the resources of the InP in a way that the number of admitted services, meeting their SLAs, is maximized. Most of the previous studies focused on performing service placement without considering the SLA of the service. Recently, SLA-aware service placement has received attention for NFV-based networks. The reliability of the services is one of the most critical requirements in the next generation of telecommunication network, especially 5G and beyond. The three service types of 5G named enhanced mobile broadband (eMBB), ultra-reliable and low-latency communication (URLLC), and machine-type communications (MTC) have specific reliability requirements \cite{osseiran2014scenarios}. As a consequence, the reliability-aware service placement for NFV-enabled NO needs to be extensively studied.

The reliability of a service in NFV depends on the reliability of the commodity servers running the VNFs of the service. In fact, a service is in a perfect running state, if all  VNFs of the service are running without failure. On the other hand, the servers of InPs can have different reliability levels, which makes the reliability-aware service placement more complicated. One common solution for providing the reliability requirement is using hot backups. It is referred to allocation of additional servers to some of the constituent VNFs of the service's SFC. The first server is called the main server, and the next servers are called the backup servers. There are two general approaches for performing backup allocation. In the first approach, backup server allocation is done after performing the main server assignment. In the second approach, the main and backup server allocations are done simultaneously. Most of the proposed methods for reliability-aware service placement is based on the first approach, which is not the optimal solution.

Generally, the most unperceived aspect of reliability-aware service placement is the dynamic nature of the service arrival and departure. More precisely, most of the previous studies have not considered the dynamic characteristic of the services. The dynamic nature of the services can dramatically affect the performance of the proposed methods. To the best of our knowledge, there is no comprehensive work for dynamic reliability-aware service placement where the main and backup servers allocation is carried out simultaneously.
The main contributions of this paper can be summarized as follows:
\begin{itemize}
\item We consider a scenario in which an NO aims to provide different service types using NFV for the users. Each service type is characterized by its arrival rate, departure rate, reliability requirements, and SFC specifications.
\item We formulate an optimization problem for dynamic reliability-aware service placement, which considers the main and backup servers allocation in one step with the objective of minimizing placement cost and maximizing the number of admitted services meeting their reliability.
\item Then, we formulate the dynamic reliability-aware service placement as an infinite horizon Markov decision process (MDP) problem to minimize the placement cost and maximize the number of admitted services. We define the state set of the MDP based on the number of active services and the number of incoming services of each type and the action set using the number of possible admitted services of each type.
\item We propose a method to estimate the idle resources of the InPs according to the number of active services in the process of finding the optimal policy of the MDP. For evaluating each possible action, we use a sub-optimal method based on the Viterbi algorithm named Viterbi-based Reliable Static Service Placement (VRSSP) algorithm.
\item We adopt the value iteration algorithm named VRSSP-based Value Iteration (VVI) to find the optimal policy of the proposed MDP based on VRSSP. During VVI, we determine the best possible arrangement of the admitted services for placement.
\item Finally, we evaluate the performance of the proposed MDP model for dynamic reliability-aware service placement. We compare the performance of the MDP model with the baseline static methods for reliability-aware service placement.
\end{itemize}

The rest of the paper is organized as follows. The existing methods for reliability-aware service placement in NFV are
reviewed in Section \ref{Section_RelatedWorks}. We introduce a dynamic reliability-aware service placement in Section \ref{Section_SystemModel}. Then, we propose an MDP model for dynamic reliability-aware service placement in Section \ref{MDP_Section}. We present the algorithm for service placement named VRSSP algorithm in Section \ref{StaticAlgorithm_Section}. In the following, we present VVI algorithm for finding the optimal policy of the introduced MDP in Section \ref{OptimalPolicyComputing_Section}. Finally, we numerically evaluate the proposed scenario for dynamic reliability-aware service placement in Section \ref{NumericalResult_Section}.
\vspace{-4mm}
\section{Related~Works}
\label{Section_RelatedWorks}
In this section, we first review the static service placement problem and then introduce the conducted research in dynamic service placement.

{\color{black}In \cite{pham2017traffic, bari2016orchestrating, mechtri2016scalable}, the static service placement problem is investigated for minimizing the placement cost by considering different cost components including server cost, cost of traffic routing, a penalty for resource fragmentation, and delay cost.
In \cite{herker2015data, fan2015grep, ding2017enhancing, fan2018framework, qu2017reliability, sun2018reliability}, the static reliability-aware service placement is considered.
In \cite{herker2015data, fan2015grep, ding2017enhancing}, the backup server allocation apart from the main server placement is done. In \cite{herker2015data, fan2015grep}, the VNF selection for backup is performed independent of backup placement. In \cite{ding2017enhancing}, an iterative cost-effective redundancy algorithm named CERA is proposed in which VNF selection for backup and backup placement are combined. The authors in \cite{fan2018framework} first map each service with the estimated number of backups, and then, provide the required reliability by adding more backups. The authors in \cite{qu2017reliability} investigate iterative backup selection with a routing procedure and endeavored to maximize link utilization while providing required reliability and delay. In \cite{sun2018reliability}, an algorithm named ensure reliability cost-saving (ER-CA) for reducing the cost of placement is presented. In \cite{kanizo2017optimizing} and \cite{kanizo2018designing}, virtual backup allocation to recover the failing middleboxes is considered. In \cite{kanizo2017optimizing}, the idea of shared backup in which each backup server is a backup for the multiple middleboxes is introduced. In \cite{kanizo2018designing}, a novel graph-based presentation for backup server allocation is proposed. In \cite{rottenstreich2016minimizing}, the idea of pipeline sharing for decreasing the number of requirement cores in NFV is investigated in a way that the average delay is minimized.}

{\color{black}In \cite{chen2019automated, chen2018multi, woldeyohannes2018cluspr, yang2018cost, cao2019dynamic, jia2018online}, service placement by considering the dynamic characteristic of the services in NFV-enabled NO is investigated. In \cite{chen2019automated, chen2018multi}, a distributed approach for dynamic service placement is proposed. In \cite{woldeyohannes2018cluspr}, an online algorithm by dynamic adjusting the number of virtual network function instances (VNFIs) is introduced. In \cite{yang2018cost}, the dynamic service placement for NO with the capability of mobile edge computing (MEC) is considered. In \cite{cao2019dynamic}, the jointly dynamic service placement and scheduling problem are modeled as a mixed-integer linear programming (MILP) by providing guaranteed quality of service (QoS). An online algorithm based on the regularization approach is proposed in \cite{jia2018online}. In \cite{eramo2017approach, eramo2017migration, liu2017dynamic}, the dynamic migration-based service placement model with considering the negative effect of migration on the QoS is considered. In \cite{khezri2018deep}, for dynamic reliability-aware service placement only using the main server, a deep reinforcement learning (Deep-RL) method is proposed. To the best of our knowledge, there is no comprehensive work for dynamic reliability-aware service placement by the simultaneous allocation of main and backup servers for NFV-enabled NO.}

\vspace{-3mm}
\section{System~Model}
\label{Section_SystemModel}

In this section, we introduce the dynamic reliability-aware service placement problem. We consider a scenario in which an NO needs to deliver services using NFV. However, the NO would use the resources of existing InPs for the placement of the incoming services. There are multiple InPs with commodity servers providing service to NOs. Each InP has several servers with different amount of resources and a certain level of reliability.
\vspace{-5mm}
\subsection{Infrastructure Network Providers (InPs)}
\label{InP_SubSec}
Let $I$ denote the set of existing InPs that NO can use their servers. We model the entire network of InPs as a uni-directed graph $G = (G^s, G^b)$, where $G^s$ is the set of servers and $G^b$ is the set of the links between the servers which can be written as
\vspace{-2mm}
\begin{align}
G^s &=\Big\{G^{s}_{i}\mid s \in \{1,2,\ldots,|S_i|\}, i \in \{1,2,\ldots,|I|\}\Big\}, \label{Server_Set} \\
G^b &=\Big\{G^{s_1,s_2}_{i_1,i_2}\mid s_1 \in \{1,2,\ldots,|S_{i_1}|\}, \; s_2 \in \{1,2,\ldots,|S_{i_2}|\}, \; i_1, i_2 \in \{1,2,\ldots,|I|\}\Big\},
\end{align}
\noindent where $G^s_i$ is the $s^\text{th}$ server of the $i^\text{th}$ InP and $G^{s_1,s_2}_{i_1,i_2}$ indicates the link between the $s_1^\text{th}$ server of the $i_1^\text{th}$ InP and the $s_2^\text{th}$ server of $i_2^\text{th}$ InP. Each server have $|R|$ resource types where the amount of $j^\text{th}$ resource type on the $s^\text{th}$ server of $i^\text{th}$ InP is denoted by $R_{i,j}^s$. Examples of such resource types are CPU, RAM, and storage. The bandwidth capacity and the unit cost of using link $G^{s_1,s_2}_{i_1,i_2}$ are denoted by $B_{i_1,i_2}^{s_1,s_2}$ and $C_{i_1,i_2}^{s_1,s_2}$, respectively. The unit cost of using the $j^\text{th}$ resource type of $i^\text{th}$ InP's server is denoted by $C_{i,j}$. Let $v_i$ indicate the failure probability of the servers of the $i^\text{th}$ InP. We assume that decreasing the failure probability marginally close to zero exponentially increases the cost of servers. As a result, the value of $C_{i,j}$ can be written as
\vspace{-3mm}
\begin{align}
C_{i,j} = \alpha_{j} e^{\beta(v_{\text{Base}}-v_i)}, \ i=1,\ldots,|I| \label{ServerCostvsReliability},
\end{align}
\noindent where $\alpha_{j}$ ($0 \leq \alpha_{j} \leq 1$) is a coefficient which indicates the importance of $j^\text{th}$ resource type for the servers of $i^\text{th}$ InP. $\beta$ is the design parameter and $v_{\text{Base}}$ is the highest acceptable failure probability. The failure probability of the servers of each InP should be lower than this threshold.
{\color{black}We know that with decreasing the failure probability of a server, the involved hardware in the server will be more expensive, which leads to an increment of server cost. The downtime of the server is one of the best metrics for determining the server cost. For example, for the servers with failure probabilities of $\{0.05, 0.03, 0.01\}$, the downtimes are $\{18.26, 10.96, 3.65\}$ days per year. Assume that the cost of the first server is $1$, we can consider the costs of the second and third servers to be $1.66$ and $5$ according to their lower downtimes.
These values of servers' cost can be obtained using the introduced exponential model. By using this model, the high-reliable servers become expensive, which leads to efficient usage of the InPs' resources by the NO.}
\vspace{-6mm}
\subsection{Characteristics of Service Requests}
\label{Characteristics_Service_Part}
For service arrival and departure, we assume that the time is divided into equal slots.
{\color{black}The concept of the slotted time is introduced to define the time evolution of the system. The length of the slot is the input of the problem and can be set to any value. For example, it can be selected based on the minimum time between two consecutive arrivals and the minimum time between two consecutive departures.
}
{\color{black}To avoid service placement for each arrival service, we consider service placement problem at the beginning of the $n^{\text{th}}$ slot for arriving services during the $(n-1)^{\text{th}}$ slot. Therefore, the service placement is performed for arrival services during a slot, which can be lead to a better result especially in terms of admission ratio compared to performing service placement problem for each incoming service.}
In 5G networks, each service has a specific service type. Examples of such types are eMBB, URLLC, and MTC. In our model, each incoming service has a specific type with certain characteristics. Let $\Upsilon_n^k \in \Upsilon$ for $k=1,\ldots,K_n$ indicate the service type for the $k^{\text{th}}$ incoming service of $n^{\text{th}}$ slot and $K_n$ is the number of incoming services in the $n^{\text{th}}$ slot. $\Upsilon$ is the set of service types defined as $\Upsilon \; = \Big\{\Upsilon^1, \Upsilon^2, \ldots, \Upsilon^{L}\Big \}$, where $L$ is the total number of the service types and $\Upsilon^l$ denotes the characteristics of $l^{\text{th}}$ service type, defined as
\vspace{-2mm}
\begin{align}
&\Upsilon^l = \Big\{F^l, d^l, b^l, U^l, f^l, \lambda^l_{\text{max}}, r^l_{u,j}, t^l_u \Big\} \label{ServiceTypeInfo}, \; 1 \leq l \leq L, \; 1 \leq j \leq |R|, \; 1 \leq t^l_u \leq |T|,
\end{align}
where $F^l$ is maximum tolerable failure reliability, $b^l$ is the required bandwidth, $U^l$ is the number of the VNFs of the $l^{\text{th}}$ service type's SFC and $r^l_{u,j}$ is the amount of the $j^{\text{th}}$ resource type required for the $u^{\text{th}}$ VNF in the SFC of the $l^{\text{th}}$ service type. $t^l_u$ indicates the VNF type of $u^{\text{th}}$ VNF and $|T|$ is the total number of VNF types.
{\color{black}Also, $d^l$ indicates the departure probability of the $l^{\text{th}}$ service type at the end of each slot. Each admitted service will remain in the system for a random number of slots and leaves the system by the end of each slot with a probability of $d^l$. The departure probability of each active service at the end of each slot is constant and independent of the number of slots that the service is active. Therefore, the service duration and the number of active services at the beginning of each slot will be memoryless. As a consequence, the number of active services at the beginning of the $n^{\text{th}}$ slot only depends on the number of active services at the beginning of the $(n-1)^{\text{th}}$ slot and the number of admitted services at the beginning of the $n^{\text{th}}$ slot.}
{\color{black}The value of $d^l$ is the input of the problem and can be set to any value.}

Finally, $f^l$ is the probability distribution function (PDF) of the number of incoming services with type $l$ in each slot defined as $f^l(m)=\Pr\{\lambda_n^l = m\}$ for $0 \leq \lambda_n^l \leq \lambda^l_{\text{max}}$,
where $\lambda_n^l$ is the number of incoming services with type $l$ in the $n^{\text{th}}$ slot and $\lambda^l_{\text{max}}$ is the maximum number of incoming services of $l^{\text{th}}$ service type. The total number of the incoming services in the $n^{\text{th}}$ slot can be written as $K_n = \sum_{l=1}^{L}{\lambda_n^l}$. The number of incoming services in each slot is independent of the incoming services in the previous slots and the number of the active services.
{\color{black}It is worth noting that a popular scenario for modeling the Internet traffic dynamics of the users in cellular networks is cycle-stationary traffic \cite{eramo2017approach, shafiq2011characterizing}. In this model, it is assumed that the traffic volume is changed periodically among $N$ intervals. For example, in \cite{eramo2017approach}, a cycle-stationary traffic scenario with $N = 24$ which is a typical value for daily traffic, is used. However, the considered model for traffic arrival and departure in our paper is more general than the mentioned cycle-stationary traffic scenario. It should be mentioned that by considering the cycle-stationary traffic scenario, the Markov characteristic for the number of the active services is preserved. Therefore, the MDP approach can also be applied to the cycle-stationary traffic scenario.
}
\vspace{-5mm}
\subsection{Service~Placement~Cost}
\label{Cost_Function}
The service placement in NFV has two major components to contribute to placement cost, namely the cost of using the servers and the cost of traffic forwarding between servers. However, other cost resources including a deployment cost for the different types of VNFs, and a penalty for violating the SLA of the incoming services can be also included.
Here, we consider three cost components to contribute to placement cost.
For this purpose, assume that $x^{l,k,s}_{n,u,i} \in \{0, 1\}$ indicates the binary decision variable for placing the $u^{\text{th}}$ VNF of $k^{\text{th}}$ incoming service of $l^{\text{th}}$ type in the $n^{\text{th}}$ slot, in $G_i^s$, where $1 \leq l \leq L$, $1 \leq u \leq U^l$, $1 \leq k \leq \lambda^l_n$, $1 \leq s \leq |S_i|$, $1 \leq i \leq |I|$ and $1 \leq n \leq \infty$. Also, $y^{l,k}_{n,u^{\prime}}(s_1,s_2,i_1,i_2) \in \{0, 1\}$ is the binary decision variable for forwarding the traffic between the $(u^{\prime})^{\text{th}}$ and $(u^{\prime}+1)^{\text{th}}$ VNF of this service, using $G^{s_1,s_2}_{i_1,i_2}$. Now, the components of placement cost for the $k^{\text{th}}$ incoming service of $l^{\text{th}}$ type in the $n^{\text{th}}$ slot can be written as follows:
\begin{enumerate}
\item \textit{Server Cost:} Let $\xi_{n,s}^{l,k}$ indicate the server cost for the $k^{\text{th}}$ incoming service of $l^{\text{th}}$ type in the $n^{\text{th}}$ slot. We assume that the cost of using a server is proportional to the amount of resources being used. As a result, the server cost for a service can expressed as
\vspace{-3mm}
\begin{align}
\textstyle \xi_{n,s}^{l,k} = {\sum_{u=1}^{U^l}{\sum_{i=1}^{|I|}\sum_{s=1}^{|S_i|} \sum_{j=1}^{|R|} x^{l,k,s}_{n,u,i} \times r^l_{u,j}\times C_{i,j}}}. \label{ServerCost}
\end{align}
\item \textit{Traffic Forwarding Cost:}  Let $\xi_{n,b}^{l,k}$ indicate the traffic forwarding cost for the $k^{\text{th}}$ incoming service of $l^{\text{th}}$ type in the $n^{\text{th}}$ slot which can be expressed as
\vspace{-3mm}
\begin{align}
\textstyle \xi_{n,b}^{l,k} =  {\sum_{u^{\prime}=1}^{U^l-1}{\sum_{i_1=1}^{|I|}\sum_{s_1=1}^{|S_{i_1}|}\sum_{i_2=1}^{|I|}\sum_{s_2=1}^{|S_{i_2}|} y^{l,k}_{n,u^{\prime}}(s_1,s_2,i_1,i_2) \times b^l\times C_{i_1,i_2}^{s_1,s_2}}}. \label{LinkCost}
\end{align}
\item \textit{VNF Deployment Cost:} We define the VNF deployment cost for the $k^{\text{th}}$ incoming service of $l^{\text{th}}$ type in the $n^{\text{th}}$ slot with $\xi_{n,d}^{l,k}$ which depends on the number of VNFs in the service's SFC and the types of VNFs. This cost can be expressed as
\vspace{-3mm}
\begin{align}
\textstyle \xi_{n,d}^{l,k} = {\sum_{u=1}^{U^l}{\sum_{i=1}^{|I|}\sum_{s=1}^{|S_i|}  x^{l,k,s}_{n,u,i} \times DC_{i,t_u^l}}}, \label{DeploymentCost}
\end{align}
\noindent where $DC_{i,t_u^l}$ indicates the deployment cost of the $(t_u^l)^{\text{th}}$ VNF type in the servers of $i^{\text{th}}$ InP.
\end{enumerate}

The placement cost for the $k^{\text{th}}$ service of $l^{\text{th}}$ type in the $n^{\text{th}}$ slot can be computed as
\vspace{-3mm}
\begin{align}
\xi_{n,p}^{l,k}=\xi_{n,s}^{l,k} + \xi_{n,b}^{l,k} + \xi_{n,d}^{l,k}. \label{Objective_Function}
\end{align}
{\color{black}According to the definition of the placement cost, in most scenarios, the placement cost of the services is heightened by increasing the number of VNFs in the service's SFC. Therefore, admitting the services with a low number of VNFs is more profitable for NO, which is not an appropriate service admission policy. For preventing this shortcoming, we consider different values for the reward of admitting different service types, $q^l$, which is introduced in Section \ref{RewardSet_Subsection}. Intuitively, the reward of admitting services should be heightened by increasing the number of VNFs in the service's SFC. However, by these values of service admitting reward, the total number of admitted services is decreased, but there would be a compromise for admitting the services with the different number of the VNFs.
}
\vspace{-6mm}
\subsection{Placement~Constraints}
{\color{black}We consider five constraints for the service placement. The first one is introduced to guarantee the allocation of the main server to each VNF and considering the possibility of backup server allocation to the respective VNF. The second and third constraints are used to prevent the violation of each server's resources and the bandwidth of each link, respectively. The fourth constraint is introduced to guarantee the allocation of an appropriate link for forwarding the traffic between the consecutive VNFs, considering the allocated servers to the VNFs. The last constraint is introduced to guarantee that the reliability requirement of each service is provided. In this way, the reliability of each service is computed as a function of the binary decision variable, $x_{n,u,i}^{l,k,s}$.}

First, we introduce constraint of the main and backup servers allocation to each VNF, which is indicated by $H_p$. This constraint is to ensure that each VNF of the incoming services is placed in one server as the main server. Also, this constraint provides the possibility of using a backup server to meet the reliability requirement. This constraint can be written as
\vspace{-3mm}
\begin{align}
\textstyle H_p: & \textstyle  1 \leq \sum_{i=1}^{|I|}{\sum_{s=1}^{|S_i|}{{x^{l,k,s}_{n,u,i} }}} \leq 2,\label{Placement_Constraint} \; 1 \leq u \leq |U^l|, \;\; 1 \leq k \leq \lambda_n^l, \;\; 1 \leq l \leq L.
\end{align}

{\color{black}According to this constraint, if a VNF has a backup server in addition to the main server, the main and backup servers are placed in different physical servers. Therefore, the main and backup servers of a VNF are physically separated.
}

Now, we introduce the constraint for the resources of the servers in each slot which is indicated by $H_g$. Let $\omega_{n,i,j}^{s}$ indicate the amount of idle resources of $j^\text{th}$ resource type in the $s^\text{th}$ server of $i^\text{th}$ InP at the beginning of $n^\text{th}$ slot $\big(0 \leq \omega_{n,i,j}^{s} \leq R^s_{i,j}\big)$. For the placement of the incoming services in this slot, NO considers the constraints on the resources of the servers as
\vspace{-3mm}
\begin{align}
H_g: &\textstyle \sum_{l=1}^{L}{\sum_{k=1}^{\lambda_n^l}{\sum_{u=1}^{U^l}{x^{l,k,s}_{n,u,i} \times r^l_{u,j}}}} \leq \omega_{n,i,j}^{s}, \label{Resource_Constraint} \; 1 \leq j \leq |R|, \; 1 \leq s \leq |S_i|, \; 1 \leq i \leq |I|.
\end{align}

The third constraint is defined for the bandwidth limitation of the connection links between the servers which is indicated by $H_b$. Assume that
$\omega_{n,i_1,i_2}^{s_1,s_2}$ indicates the amount of remaining bandwidth in the connection link between the $s_1^\text{th}$ server of $i_1^\text{th}$ InP and the $s_2^\text{th}$ server of $i_2^\text{th}$ InP at the beginning of $n^\text{th}$ slot $\big(0 \leq \omega_{n,i_1,i_2}^{s_1,s_2} \leq R_{i_1,i_2}^{s_1,s_2}\big)$. For the placement of the incoming services in each slot, NO will consider the constraints on the bandwidth of the links as
\vspace{-3mm}
\begin{align}
\textstyle H_b: &\textstyle  \sum_{l=1}^{L}{\sum_{k=1}^{\lambda_n^l}{\sum_{u^{\prime}=1}^{U^l-1}{y^{l,k}_{n,u^{\prime}}(s_1,s_2,i_1,i_2) \times b^l}}} \leq \omega_{n,i_1,i_2}^{s_1,s_2}, \label{Bandwidth_Constraint} \\
&\textstyle  1 \leq s_1 \leq |S_{i_1}|, \; 1 \leq s_2 \leq |S_{i_2}|, \; 1 \leq i_1,i_2 \leq |I|. \notag
\end{align}

The fourth constraint is defined for traffic forwarding of each service indicated by $H_f$. For the placement of the incoming services in each slot, NO will consider the following constraints.
\vspace{-10mm}
\begin{align}
H_f:&y^{l,k}_{n,u^{\prime}}(s_1,s_2,i_1,i_2) \leq x^{l,k,s_1}_{n,u^{\prime},i_1}, \;  y^{l,k}_{n,u^{\prime}}(s_1,s_2,i_1,i_2) \leq x^{l,k,s_2}_{n,u^{\prime}+1,i_2}, \label{Forwarding_Constraint} \\
&y^{l,k}_{n,u^{\prime}}(s_1,s_2,i_1,i_2) \geq \big (x^{l,k,s_1}_{n,u^{\prime},i_1} + x^{l,k,s_2}_{n,u^{\prime}+1,i_2} - 1 \big), \; \; 1 \leq l \leq L, \;\; 1 \leq k \leq \lambda^l_n, \; 1 \leq u^{\prime} \leq U^l -1,  \notag \\
& 1 \leq s_1 \leq |S_{i_1}|, \; 1 \leq s_2 \leq |S_{i_2}|, \; 1 \leq s \leq |S_i|, \; 1 \leq i,i_1,i_2 \leq |I|. \notag
\end{align}

This constraint implies that if $u^{\prime}$ and $u^{\prime}+1$ VNFs of a service are placed in servers $G_{i_1}^{s_1}$ and $G_{i_1}^{s_1}$, respectively, the traffic between these VNFs should be routed using link $G_{i_1,i_2}^{s_1,s_2}$.

The final constraint is defined for the reliability requirement of the incoming services.
We indicate the failure probability of $k^{\text{th}}$ incoming service of $l^{\text{th}}$ type in the $n^{\text{th}}$ slot with $f_n^{l,k}$. To obtain $f_n^{l,k}$, we calculate the probability of being in the running state (i.e., not being failed) which is indicated by $p_n^{l,k}$. We know that a service is in running state if none of the VNFs of that service fails. As a result, we should determine the failure probability of the VNF as a function of binary decision variable, $x^{l,k,s}_{n,u,i}$. Let $f_{n,u}^{l,k}$ denote the failure probability of $u^{\text{th}}$ VNF of $k^{\text{th}}$ service of $l^{\text{th}}$ type in the $n^{\text{th}}$ slot, which can be computed as $
\textstyle f_{n,u}^{l,k} = \prod_{i=1}^{|I|}\Big(\prod_{s=1}^{|S_i|} \rho_{n,u,i}^{l,k,s}\Big),  
$
where $\rho_{n,u,i}^{l,k,s}=v_i$ when $x_{n,u,i}^{l,k,s}=1$, and otherwise it is 1 ($v_i$ is the failure probability of $i^{\text{th}}$ InP).
Now, we can compute the probability of being in the running state and failure probability of the placement for the $k^{\text{th}}$ service of $l^{\text{th}}$ type in the $n^{\text{th}}$ slot as
$p_n^{l,k} = \prod_{u=1}^{U_l}{(1-f_{n,u}^{l,k})}, \; f_n^{l,k} = 1-p_n^{l,k}$, and
\label{eq:Sevice_Correct_Probability}
the reliability constraint can be written as
\vspace{-4mm}
\begin{align}
\textstyle H_r: &1- \textstyle \prod_{u=1}^{U_l}{\Big(1-\prod_{i=1}^{|I|}\big(\prod_{s=1}^{|S_i|} \rho_{n,u,i}^{l,k,s}\big)\Big)} \leq F^l
\label{Reliability_Constraint}, \quad 1 \leq k \leq \lambda_n^l, \; 1 \leq l \leq L.
\end{align}

\vspace{-8mm}
\subsection{Dynamic Reliability-Aware Service Placement Problem}
\label{OptimizationProblemPart}
The optimization problem of dynamic reliability-aware service placement is
\vspace{-4mm}
\begin{align}
\textstyle \min_{x^{l,k,m}_{n,u,i}, y^{l,k}_{n,u^{\prime}}(s_1,s_2,i_1,i_2)} & \textstyle \sum_{n=1}^{\infty}{\sum_{l=1}^{L} { \sum_{k=1}^{\lambda_n^l}{\xi_{n,p}^{l,k}} }}\label{eq:Objective_11}\\
\text{subject to} \;\; & H_p, H_g, H_b, H_f,  H_r.
\end{align}

{\color{black}This optimization problem aims to minimize the total placement cost during the infinite time for all incoming services of all service types. The considered constraints are for main and backup server allocation, for the resource capacity of the servers, for the bandwidth capacity of the links, for the traffic routing, and for the reliability requirement of the services which are indicated by $H_p$, $H_g$, $H_b$, $H_f$, and $H_r$, respectively. The optimization variables are $x^{l,k,m}_{n,u,i}$ and $y^{l,k}_{n,u^{\prime}}(s_1,s_2,i_1,i_2)$,  where  $x^{l,k,m}_{n,u,i}$ is the binary decision variable for placing the $u^{\text{th}}$ VNF of $k^{\text{th}}$ incoming service of $l^{\text{th}}$ type in the $n^{\text{th}}$ slot, using server $G_{i}^{s}$, and $y^{l,k}_{n,u^{\prime}}(s_1,s_2,i_1,i_2)$ is the binary decision variable for routing the traffic between the $(u^{\prime})^{\text{th}}$ and $(u^{\prime}+1)^{\text{th}}$ VNFs of $k^{\text{th}}$ incoming service of $l^{\text{th}}$ type in the $n^{\text{th}}$ slot, using link $G_{i_1,i_2}^{s_1,s_2}$.}

This optimization problem cannot be solved with standard static optimization techniques due to the dynamic nature of the parameters. More precisely, the amount of idle resources in each InP depends on the number of active services.
{\color{black}NO should take into consideration the number of active service for admitting new services. In such a scenario, the dynamic programming (DP) techniques can be beneficial. An MDP, as a DP technique, is a model of an agent interacting synchronously with a world. The agent receives as input the state of the world and takes as output actions, which affect the state of the world. The most important character for modeling a dynamic optimization problem with MDP is the memoryless characteristic of the state. According to some aforementioned assumption, the number of the active services, which can be considered as the state of the problem, is memoryless. Then, for responding to the characteristics of the dynamic reliability-aware service placement, MDP can be an appropriate idea.} {\color{black} Table \ref{table1} includes the notations used in the problem and proposed MDP model.}

{\color{black}
\begin{table*}
	\caption{\small{Notation table including all notations of optimization problem and MDP model}}
	\label{table1}
	\vspace{-3mm}
	\begin{minipage}[t]{0.5\textwidth}
		\begingroup
		\begin{tabular}[t]{>{\centering\arraybackslash}p{0.18\textwidth}p{0.73\textwidth}}
			\hline
			
			 {\bf Symbol} & {\bf Description}\\
			
			\hline

			$G^s, G^b$ &Set of servers and links \\
			$G^s_i$ &The $s^{\text{th}}$ server of $i^{\text{th}}$ InP.\\
			$G^{s_1,s_2}_{i_1,i_2}$ &The link between $(s_1)^{\text{th}}$ server of $(i_1)^{\text{th}}$ InP and $(s_2)^{\text{th}}$ server of $(i_2)^{\text{th}}$ InP.\\
			$R^s_{i,j}$ &The amount of $j^{\text{th}}$ resource type in $G^s_i$\\
			$B_{i_1,i_2}^{s_1,s_2}, C_{i_1,i_2}^{s_1,s_2}$ &The cost and bandwidth of $G^{s_1,s_2}_{i_1,i_2}$\\
			$C_{i,j}, v_i$ &The cost of using $j^{\text{th}}$ resource type and failure probability of servers of $i^{\text{th}}$ InP\\
			$F^l, b^l, U^l$  &Failure probability, bandwidth requirement, and number of VNFs in the $l^{\text{th}}$ service type.\\
			$d^l, f^l, \lambda^l_{\text{max}}$  &Departure probability, PDF of arrival services, maximum number of arrivals of $l^{\text{th}}$ service type.\\
			$r_{u,j}^l, t_u^l$ &Resource requirement of $j^{\text{th}}$ type and VNF type of $u^{\text{th}}$ VNF of $l^{\text{th}}$ service type \\
			$x_{n,u,i}^{l,k,s}$ &Decision varibable of placing $u^{\text{th}}$ VNF of $k^{\text{th}}$ service of $l^{\text{th}}$ type in the $n^{\text{th}}$ slot, in $G^s_i$\\
			$\xi_{n,s}^{l,k}, \xi_{n,b}^{l,k}, \xi_{n,d}^{l,k}$ &Server cost, traffic forwarding cost, deployment cost of $k^{\text{th}}$ service of $l^{\text{th}}$ type in the $n^{\text{th}}$ slot\\
			$\xi_{n,p}^{l,k}$ &Placement cost of $k^{\text{th}}$ service of $l^{\text{th}}$ type in $n^{\text{th}}$ slot\\
			$\omega_{n,i,j}^s$ &Remaining resource of $j^{\text{th}}$ type of $G^s_i$ in $n^{\text{th}}$ slot\\
			$\omega_{n,i_1,i_2}^{s_1,s_2}$ &Remaining bandwidth of $G^{s_1,s_2}_{i_1,i_2}$ in $n^{\text{th}}$ slot.\\
			$p_n^{l,k}, f_n^{l,k}$ &Running state probability and failure probability of $k^{\text{th}}$ service of $l^{\text{th}}$ type in the $n^{\text{th}}$ slot\\
			$\Omega_S^R, \Omega_S^D, \Omega_S^I$ &State set of idle resource, active service, and incoming service \\
			$\lambda_n, \sigma_n$ &State of incoming service and active service in $n^{\text{th}}$ slot\\

		\end{tabular} \endgroup
	\end{minipage} \hfill \vline
	\begin{minipage}[t]{0.50\textwidth}
		\begingroup
		\begin{tabular}[t]{>{\centering\arraybackslash}p{0.14\textwidth}p{0.73\textwidth}}
			
			\hline
			 {\bf Symbol} & {\bf Description}\\	

			\hline
			$\lambda_n^l, \sigma_n^l$ &State of incoming service and active service of $l^{\text{th}}$ type in $n^{\text{th}}$ slot\\
			$\sigma_{\text{max}}^l$ &Maximum number of active services of $l^{\text{th}}$ type\\
			$|\Omega_S^D|, |\Omega_S^I|$ &Size of state sets of active service and incoming service.\\
			$\Omega_S, \Omega_A^f$ & State set and set of feasible action\\
			$S_n, A_n$ &State and action of system in the $n^{\text{th}}$ slot\\
			$P(S,A,S^\prime)$ &Probability of ending in state $S^\prime$, if agent take action $A$ in state $S$ \\
			$\mathbf{P}^i$ &Transition matrix over $\Omega_S^D$ when state of incoming service is $i$\\
			$\Delta(S,A)$ &Validation metric of taking action $A$ in state $S$\\
			$R(S,A)$ &Reward of taking action $A$ in state $S$\\
			$e^{l,k}, \xi^{l,k}_p$ &Failure probability and placement cost of $k^{\text{th}}$ service of $l^{\text{th}}$ type which are results of VRSSP algorithm\\
			$\rho$ &Placement arrangement (input of VRSSP algorithm)\\
			$N_{i,j}^s[l,k]$ &Usage of $j^{\text{th}}$ resource type of $G_i^s$ for placement of $k^{\text{th}}$ service of $l^{\text{th}}$ type (output of VRSSP algorithm)\\
			$\omega_{i,j}^{s,\eta_S}$ &Idle resources of $j^{\text{th}}$ type of $G_i^s$ when state is $S$\\
			$\alpha^{\eta_S}$ &Updating factor of estimating idle resource for state $S$\\
			$N_{i,j}^s(S,A)$ &Resource usage of $j^{\text{th}}$ type of $G_i^s$ for taking action $A$ in state $S$\\
			$R_{\text{opt}}^{S,A}, \rho_{\text{opt}}^{S,A}$ &Reward of optimum arrangement and optimum arrangement for taking action $A$ in state $S$\\

		\end{tabular}
		\endgroup
		
	\end{minipage}
	\hrule width 1.0\textwidth
	\vspace{-10mm}
\end{table*}
}
\vspace{-4mm}
\section{AN~MDP~MODEL~OF~DYNAMIC~RELIABILITY-Aware~Service~PLACEMENT}
\label{MDP_Section}
As we know, MDP provides a mathematical framework for decision making in problems where outcomes are partly random and partly depend on the actions of the decision maker. It should be mentioned that NO is the agent making a decision by formulating and solving the MDP.
An MDP problem is characterized by a four-tuple $(\Omega_S, \Omega_A, \Omega_P, \Omega_R)$ where $\Omega_S$ is the state set, $\Omega_A$ is the action set, $\Omega_P$ is the probability set, and $\Omega_R$ is the reward set. In the MDP framework, it is assumed that, although there may be a great deal of uncertainty about the effects of an agent’s actions, there is never any uncertainty about the agent’s current state. Also, we assume that the agent has complete and perfect perceptual abilities.
For modeling the dynamic reliability-aware service placement problem with MDP, we define these sets in the following.
\vspace{-6mm}
\subsection{Set~of~States}
\label{State_Set_Part}
The definition of an appropriate state set is the most important parts of modeling a problem as an MDP. The state of the system has three components. The first one is the state of the idle resources of servers in different InPs, at the beginning of each slot, which is indicated by $\Omega^R_S$. The second component is the state of the active services, which is equal to the number of active services of each type at the beginning of each slot indicated by $\Omega^D_S$. We know that active services at the beginning of each slot are the services that are admitted in the previous slots and have not left the network. The last component of the state set is the state of the incoming services at the beginning of each slot, which is indicated by $\Omega^I_S$. In the following, we define the components of the state set.

\begin{enumerate}
\item $\Omega^R_S$: The state set of the idle resources of the InPs can be written as
\vspace{-4mm}
\begin{align}
&\Omega^R_S = \Big\{(\omega_{i,j}^{s}) \, \big | \, 0 \leq \omega_{i,j}^{s} \leq R_{i,j}^s \Big\}, \label{RemainResourceState} \; 0 \leq s \leq |S_i|, \; 0 \leq i \leq |I|, \; 0 \leq j \leq |R|,
\end{align}
where $\omega_{i,j}^s$ indicates the state of idle resources of $j^\text{th}$ resource type in the $s^\text{th}$ server of $i^\text{th}$ InP. The state of idle resources of the servers in the $n^\text{th}$ slot can be written as $\big\{(\omega_{n,i,j}^{s})\big\} \in \Omega^R_S$. As seen in \eqref{RemainResourceState}, the state of idle resources of the servers is continuous. We know that MDPs with continuous state space has high computational complexity and can be impractical. Later on, we discuss this challenge.
\item $\Omega^D_S$: We assume that each incoming service has a specific type that determines the characteristics of the service. Therefore, the state set of the active services can be written as
\vspace{-10mm}
\begin{align}
&\Omega^D_S = \Big\{(\sigma^1, \sigma^2, \ldots, \sigma^L) \, \big | \, 0 \leq \sigma^l \leq \sigma^l_{\text{max}} \Big\} \label{ExistState},
\end{align}
where $\sigma^l$ and $\sigma^l_{\text{max}}$ are the number of active services and the maximum number of active services for the $l^\text{th}$ service type, respectively. If the number of active services for the $l^\text{th}$ service type at the beginning of a slot is $\sigma^l_{\text{max}}$, NO will not admit any new incoming services with type $l$. We indicate the number of active services with type $l$ at the beginning of $n^\text{th}$ slot with $\mathbf{\sigma}_n = (\sigma^1_n, \sigma^2_n, \ldots, \sigma^L_n)$  where $\mathbf{\sigma}_n \in \Omega^D_S$.
\item $\Omega^I_S$: We consider the state of the incoming services as the number of incoming services of each service type separately. Therefore, the state of the incoming services is
\vspace{-3mm}
\begin{align}
&\Omega^I_S = \Big\{(\lambda^1, \lambda^2, \ldots, \lambda^L) \, \big | \, 0 \leq \lambda^l \leq \lambda^l_{\text{max}} \Big\} \label{IncomingeState},
\end{align}
where $\lambda^l$ and $\lambda^l_{\text{max}}$ are the number of incoming service and the maximum number of incoming service for the $l^\text{th}$ service type, respectively. The number of incoming services in the $n^\text{th}$ slot is indicated by $\mathbf{\lambda}_n = (\lambda^1_n, \lambda^2_n, \ldots, \lambda^L_n)$  where $\mathbf{\lambda}_n \in \Omega^I_S$.
\end{enumerate}

The total state set of the system can be written as $\Omega_S = \Omega^R_S \times \Omega^D_S \times \Omega^I_S$, which is a massive state space. For example, the state space of idle resources of the InPs, $\Omega^R_S$, is continuous, and we have to use continuous state space for our MDP problem. Solving the MDP problems with continuous state space is time-consuming.
However, we know that the state of the idle resources of the servers in different InPs, at the beginning of each slot depends on the number active services at the beginning of the corresponding slot. More precisely, if the number of active services at the beginning of each slot is known, the state of the idle resources can be estimated. As a result, the state set of the proposed MDP can be considered as $\Omega_S = \Omega^D_S \times \Omega^I_S$.
For this purpose, we should introduce a method for estimating the state of the idle resources of the InPs according to the number of active services. More precisely, for each $(\sigma^1, \sigma^2, \ldots, \sigma^L) \in \Omega^D_S$, we should determine the state of idle resources, $\big\{(\omega^s_{i,j})\big\}$. We discuss more details about this method in Section \ref{OptimalPolicyComputing_Section}.
Now, we can write the total state of the system in the $n^\text{th}$ slot as ${S}_n = \{\mathbf{\lambda}_n, \mathbf{\sigma}_n\} \in \Omega_S$. We assume that the state of the incoming services and active services are independent. The size of this state set is indicated by $\big|\Omega_S\big|$ which can be computed as $\big|\Omega_S\big| =  \big|\Omega^D_S\big| \times \big|\Omega^I_S\big|$, where $\big|\Omega^D_S\big|$ and $\big|\Omega^I_S\big|$ are the size of state sets of the active services and incoming services, and can be calculated as $\big|\Omega^D_S\big| = \prod_{l=1}^{L}{(\sigma^l_{\text{max}} + 1)}, \; \big|\Omega^I_S\big| =  \prod_{l=1}^{L}{(\lambda^l_{\text{max}} + 1)}$.
{\color{black}It is worth noting that due to the random nature of the number of the incoming services in each slot and due to the independence of the number of the incoming services from the number of the active services in each slot, the inclusion of the number of incoming services in each slot in the state space is necessary to achieve the optimal policy. Otherwise, NO cannot differentiate between the different states of the number of incoming services of each type in a determined state of the active services and idle resources, which leads to a sub-optimal policy.}
\vspace{-6mm}
\subsection{Action~Set}
\label{Action_Set_Part}
The introduced decision variables in \ref{OptimizationProblemPart} are $x^{l,k,s}_{n,u,i}$ and $y^{l,k}_{n,u^{\prime}}(s_1,s_2,i_1,i_2)$ which indicate the actions for the placement of the VNFs and traffic forwarding of the services, respectively.
Therefore, we can define the action set as, $\Omega_A = \big \{ \big(x^{l,k,s}_{u,i}, y^{l,k}_{u^{\prime}}(s_1,s_2,i_1,i_2)\big) \Big| x^{l,k,s}_{u,i}, y^{l,k}_{u^{\prime}}(s_1,s_2,i_1,i_2) \in \{0,1\} \big \}$.
Even though this action set can result in optimal placement, the implementation of such action set is computationally complex and cannot be used in a practical scenario. More precisely, to determine the optimal policy in each state, all possible actions should be examined, which means a large number of exhaustive searches. Therefore, we would revise the definition of the action set. According to the definition of the state set in \ref{State_Set_Part}, the number of the possible admitted services for each service type can be considered as the action set of MDP as
\vspace{-2mm}
\begin{align}
&\Omega_A = \Big \{ (a^1, a^2, \ldots, a^L) \, \big | \, 0 \leq a^l \leq \lambda^l_{\text{max}} \Big \},  \label{Action_Set}
\end{align}
where $a^l = M$ means admitting $M$ services of $l^{\text{th}}$ service type according to the requirement of this service type. We notice that for each possible action, the placement of admitted services is not considered. Therefore, we should determine the placement of admitted services, using another static algorithm. For this purpose, we introduce Viterbi-based Reliable Static Service Placement (VRSSP) algorithm, in the next section. We consider this algorithm static because it determines the placement of services regardless of the cost in the next slots. This algorithm takes the resources of InPs and admitted services as the inputs and returns the placement of admitted services as an output. In each state, depending on the state of the active and incoming services, some of the actions in $\Omega_A$ are not feasible. As a result, the feasible actions $\Omega_A^f \subset \Omega_A$, for each state $S \in \Omega_S$ where $S = \big\{(\lambda^1,\lambda^2,\ldots,\lambda^L),(\sigma^1,\sigma^2,\ldots,\sigma^L)\big\}$, can be computed as
\vspace{-2mm}
\begin{align}
\Omega_A^f = \Big \{ (a^1, a^2, \ldots, a^L) \, \big| \, a^l + \sigma^l \leq \sigma_{\text{max}}^l, \, a^l \leq \lambda^l \Big \}. \label{FeasibleAction}
\end{align}

We know that depending on the state of the idle resources, providing the reliability requirement for some of the admitted services is not feasible. In other words, in each state, some of the actions in $\Omega_A^f$ are not feasible because of the limitation in the resources of the InPs. We should design the reward function in a way that prevents NO to choose such actions as an optimal action in each state. We completely discuss the reward function of each action in the following.
\vspace{-6mm}
\subsection{Transition~Probability}
The probability set $\Omega_P: \Omega_S \times \Omega_A \rightarrow \Omega_S$ is the state transition function, giving for each state and action, a probability distribution over the state set. Therefore, we write $P\big({S}_{n-1}, {A}_{n-1}, {S}_{n}\big)$ for the probability of ending in state ${S}_{n}$, given that the agent starts in state ${S}_{n-1}$ and takes action ${A}_{n-1}$.
In the service placement problem, the state transition depends on the current state, the taken action, the departure distribution of the active services, and the distribution of service incoming,
For defining state transition matrix $\mathbf{P}$, we indicate the probability distribution over the state space in the $n^\text{th}$ slot by $\mathbf{b}_n$, which is a vector with a length of $\big|\Omega_S\big|$ defined as
\vspace{-3mm}
\begin{align}
\mathbf{b}_n &= \big(\mathbf{b}_n^{1}, \mathbf{b}_n^{2}, \ldots, \mathbf{b}_n^{|\Omega^I_S|}),  \quad \mathbf{b}_n^i = \big({b}_n^{i, 1}, {b}_n^{i, 2}, \ldots, {b}_n^{i, |\Omega^D_S|} \big), \label{Belief_Vector} \\
{b}_n^{i, j} &= \Pr\Big(\mathbf{\lambda}_n = (i^1, i^2, \ldots, i^L), \, \mathbf{\sigma}_n = (j^1, j^2, \ldots, j^L)\Big) ,
\end{align}
where $\mathbf{b}_n^i$ is a vector that indicates the probability distribution over the state space of the active services when the state of incoming service is $i$, and $b_n^{i,j}$ indicates the probability that the state of incoming service is $i$, and the state of active service is $j$. The state of the incoming services is $i$, and the state of the active services is $j$, when we have
\vspace{-3mm}
\begin{align}
1+\sum_{l=1}^{L}{i^l \times \delta_{\lambda}^l} &= i, \; \delta_{\lambda}^l =\begin{cases} \prod_{r=1}^{l-1}{ (\lambda^r_{\text{max}}+1)}, & l \geq 2, \\ 1, & l = 1,\end{cases} \;\; 1 \leq i \leq |\Omega^I_S|, \label{activeserviceind}\\
1+\sum_{l=1}^{L}{j^l \times \delta_{\sigma}^l} &= j, \;   \delta_{\sigma}^l =\begin{cases} \prod_{r=1}^{l-1}{( \sigma^r_{\text{max}}+1)}, & l \geq 2, \\ 1, & l = 1, \end{cases} \;\; 1 \leq j \leq |\Omega^D_S|. \label{incomeserviceind}
\end{align}
{\color{black}where $\delta^l_{\sigma}$ and $\delta^l_{\lambda}$ are the auxiliary variables introduced to facilitate the computation of the state transition matrix, $\mathbf{P}$.}

We assume that the number of incoming services in the $n^{\text{th}}$ slot is independent of the incoming services and active services in the $(n-1)^{\text{th}}$ slot. Also, the number of active services in the $n^{\text{th}}$ slot depends on the state of active service and the taken action in the $(n-1)^{\text{th}}$ slot. Therefore, $S_n$ is independent of the state of incoming services in the $(n-1)^{\text{th}}$ slot, and the state transition matrix, $\mathbf{P}$, is a $|\Omega^D_S| \times |\Omega_S|$ matrix and can be written as $\mathbf{P} = \left[\begin{array}{cccc} \mathbf{P}^{1} & \mathbf{P}^{2} &\cdots & \mathbf{P}^{|\Omega^I_S|} \end{array}\right]$,
where $\mathbf{P}^{i}$ is a $|\Omega^D_S| \times |\Omega^D_S|$ matrix which is the transition matrix over the state space of the active services when the state of incoming service is $i$. Each element of $\mathbf{P}^{i}$ is defined as
\vspace{-3mm}
\begin{align}
\mathbf{P}^i(j,k) = \Pr&\Big(\mathbf{\sigma}_n = (k^1,k^2,\ldots, k^l), \mathbf{\lambda}_n = (i^1,i^2,\ldots, i^l) \, \big|  \mathbf{\sigma}_{n-1} = (j^1,j^2,\ldots, j^l)\Big),  \label{Transition_Matrix_Partial1} \\
=\Pr&\Big(\mathbf{\sigma}_n = (k^1,k^2,\ldots, k^l) \big| \mathbf{\sigma}_{n-1} = (j^1,j^2,\ldots, j^l)\Big)
\times \Pr \Big(\mathbf{\lambda}_n = (i^1,i^2,\ldots, i^l) \Big), \label{Transition_Matrix_Partial2}
\end{align}
where the state of the active services in the $(n-1)^\text{th}$ and $n^\text{th}$ slot are $j$ and $k$, respectively. The terms of being in state $i$ for the incoming services and being in state $j$ and $k$ for the active services are defined in \eqref{activeserviceind} and \eqref{incomeserviceind}. We could write \eqref{Transition_Matrix_Partial2} due to assumption that the state of the incoming services in the $n^{\text{th}}$ slot is independent of the state of the active services in the $(n-1)^{\text{th}}$ slot. The probabilities in \eqref{Transition_Matrix_Partial2} can be computed as follows:
\vspace{-2mm}
\begin{align}
\Pr&\Big(\mathbf{\sigma}_n = (k^1,\ldots, k^l) \big| \mathbf{\sigma}_{n-1} = (j^1,\ldots, j^l)\Big)
= \begin{cases}  0 & \exists{l}, \, (j^{l} - k^{l}) < 0,  \\ \prod_{l=1}^{L}{(d^{l})^{j^{l} - k^{l}}} & \text{otherwise}, \end{cases}\label{Transition_Matrix_Partial3}  \\
\Pr&\Big(\mathbf{\lambda}_n = (i^1,i^2,\ldots, i^l) \Big) = \textstyle \prod_{l=1}^{L}{f^l(i^l)} \label{Transition_Matrix_Partial4}.
\end{align}

Now, we can write the state transition probability, $P\big({S}_{n-1}, {A}_{n-1}, {S}_{n}\big)$, as
\vspace{-4mm}
\begin{align}
P\big({S}_{n-1}, {A}_{n-1}, {S}_{n}\big) &=  \mathbf{P}^i(j,k), \label{State_Transition}\\
1+\textstyle \sum_{l=1}^{L}{\lambda^l_n \times \delta_{\lambda}^l} &= i, \notag \;\; 1+\textstyle \sum_{l=1}^{L}{\sigma^l_n \times \delta_{\sigma}^l} = k, \;\; 1+\sum_{l=1}^{L}{(\sigma^l_{n-1} + a^l_{n-1}) \times \delta_{\sigma}^l} =j,\notag
\end{align}
where the values of $\delta_{\lambda}^l$ and $\delta_{\sigma}^l$ are defined in \eqref{activeserviceind} and \eqref{incomeserviceind}.
\vspace{-6mm}
\subsection{Reward~Function}
\label{RewardSet_Subsection}
In MDP, the reward set $\Omega_R: \Omega_S \times \Omega_A \rightarrow \mathbf{R}$ is the reward function, giving the expected immediate reward gained by the agent for taking each action in each state. We use $R(S, A)$ for the expected reward of taking action $A \in \Omega_A^f$ in state $S \in \Omega_S$. For the evaluation of $R(S, A)$ we define a validation metric $\Delta(A, S)$, where $\Delta(A, S) = 0$, if there is not enough resources for main server placement, and otherwise $\Delta(A, S) = 1$. As mentioned before, we define the action set as the number of possible admitted services. It was also noted that for each action, the placement of admitted services should be characterized using another static algorithm named VRSSP algorithm. In some cases for the taken action and idle resources of the InPs (estimated from the number of the active services), main server placement for some of the services is not possible because of the lack of resources in InPs. For this type of actions in such states, $\Delta(A, S) = 0$. Now, we can define the reward function regarding the reward of admitting each service, providing its reliability requirement and the placement cost of admitting each service as
\begin{align}
&R(S, A) =  \begin{cases} 0 & \Delta(A, S) = 0, \\ \sum_{l=1}^{L}{\sum_{k=1}^{a^l}{q^l\times I(F^l-e^{l,k}) - \xi_{p}^{l,k}}}  & \Delta(A, S) = 1,\end{cases} \label{RewardFunction}
\end{align}
where $q^l$ is the reward of admitting $l^{\text{th}}$ service type and $I(\cdot)$ is the indicator function where $I(x) = 1$ if $x \geq 0$, and otherwise $I(x) = 0$. Also, $e^{l,k}$ and $\xi_{p}^{l,k}$ are the failure probability and placement cost of the $k^{\text{th}}$ service of $l^{\text{th}}$ type, respectively, which are results of running the VRSSP algorithm. We explain the inputs and outputs of the VRSSP algorithm, in the next section.

\vspace{-2mm}
\section{Viterbi-based Reliable Static Service Placement (VRSSP) Algorithm}
\label{StaticAlgorithm_Section}
In this section, we present an algorithm for static service placement problem, which is named VRSSP algorithm. The introduced algorithm is called static because of solving service placement problem without considering the effect of possible placement solutions on the future slots. We indicate the input of the VRSSP algorithm as $V_I = ({A}, \mathcal{\rho}, \mathcal{O}) $. The first input, ${A}=\{a^1,a^2,\ldots,a^L\}$, determines the number of services that is  admitted from each service type.
{\color{black}The second input is defined as $\rho = (\rho_1, \rho_2, \ldots, \rho_{K})$, where $K = \sum_{l=1}^{L}{a^l}$, and $1 \leq \rho_k \leq L$ determine the placement arrangement of the services which can be admitted according to action $A$. In this way, $\rho_k = l$ means that the type of $k^{\text{th}}$ service for placement is $l$. The order of the services which should be placed by the VRSSP algorithm is effective on the output of this algorithm. For example, for a given action $A$, the order of the services which should be placed by the VRSSP algorithm can be changed for the different orders. Therefore, we consider the order of the services for placement as one of the inputs of the VRSSP algorithm. Also, different orders can lead to different placement cost. These two points directly affect the reward of each action. As a result, the optimum placement arrangement can be specified during the procedure of determining the optimal policy, which is explained in more details, in Section \ref{Arrangment_SubSection}.}
The last input determines the state of idle resources in the InPs which can be used for service placement, defined as $\mathcal{O} = \{(\omega_{i,j}^{s})\} \in \Omega^R_S$. For each input, the algorithm determines the validation metric, $\Delta(A, S)$, defined in Section \ref{RewardSet_Subsection}. In the case of valid input, the algorithm will also determine the placement cost, $\xi_{p}^{l,k}$, failure probability, $e^{l,k}$ and the amount of usage resources of servers, $N^s_{i,j}[l,k]$. Therefore, the output of the VRSSP algorithm is $V_O = (\Delta, \xi_{p}^{l,k}, e^{l,k}, N^s_{i,j}[l,k])$ where $1 \leq i \leq |I|, 1 \leq s \leq |S_i|$, $1 \leq j \leq |R|$, $1 \leq l \leq L$, and $1 \leq k \leq K$.

In the VRSSP algorithm, we apply the idea of the Viterbi algorithm for finding the most likely sequence of hidden states (called the Viterbi path) that results in a sequence of observed events. A Viterbi path is defined as an example of a sequence for observed events. The Viterbi algorithm first models the states of the problem and their transitions as a multistage graph. The number of stages is equal to the number of observed events. In each stage, one of the observed events is considered. The state of Viterbi algorithm is defined as the possible events in each stage. For the transition between all pairs of the states in consecutive stages, a transition cost is defined according to the objective function of the considered optimization problem. For each state of a stage, a path with minimum cost is selected as the survived path between the input paths to the respective state. It is worth noting that for determining the survived path of a state, the cumulative cost of input paths to the respective state is considered. Finally, in the last stage, each possible state has a survived path with a cost. The survived path of state with minimum cost is selected as a Viterbi path in the last stage. According to this description, first, we determine state, stage, and transition cost in our problem.

Let $L_V = \sum_{k=1}^K{2 U^{\rho_k}}$ denote the number of stages in the VRSSP algorithm, where the coefficient $2$ is used to consider two server assignments including the main and backup servers for each VNF. We assume that the NO can assign at most one backup server for each VNF of a service. In each stage, placement of the main or backup servers for a specific VNF is performed. For example, in the first stage, the placement of the main server for the first VNF of the first service is performed. In the second stage, the placement of the backup server for the first VNF of the first service is performed. In the $(2 \times U^{\rho_1}+1)^{\text{th}}$ stage, the placement of the main server for the first VNF of the second service is performed. We notice that $U^{\rho_1}$ is the number of VNFs for the first service. It is worth noting that in each stage, several candidates for the placement of the main or backup server for the respective VNF is determined. In the final stage, the certain placement of the main and backup server for all VNFs of all services is specified. In Appendix \ref{appendix_VRSSP}, the detailed procedure of determining the placement for the services of a specific input $V_I = ({A}, \mathcal{\rho}, \mathcal{O})$ using the idea of Viterbi algorithm is indicated.
\\
\vspace{-0.5mm}

\begin{algorithm}[!htbp]
	
	\small
		
	\caption{\small{Viterbi-based Reliable Static Service Placement (VRSSP) Algorithm}}
	
	\label{Alg1}
	\vspace{-1mm}
	\textbf{Viterbi algorithm input:} $V_I = ({A}, \mathcal{\rho}, \mathcal{O}), K=\sum_{l=1}^{L}{a^l}$,
	$L_V=\sum_{k=1}^K{2U^{\rho_k}}$, $X_0 = 0, \, \phi_0^0 = 0$, $SR_0^0[i,s,j] = \omega_{i,j}^{s}$,
	$\Delta = 1, \xi_{p}^{l,k} = 0, e^{l,k} = 1, p^{l,k} = 1, N^s_{i,j}[l,k] = 0$\\ 
	\vspace{-1mm}
	\For{($m =1:L_V$)}
	{
		\vspace{-2mm}
		Determine $k_m$ and $u_m$. \label{Alg_Service}\\
		\vspace{-2mm}
		Determine the set of states, $X_m$ using (\ref{State_Determination}) \label{Alg_State}.\\
         	\vspace{-2mm}
        	\For{$x_2 \in X_m$}
		{\label{Alg_InnerLopp_Start}
			\vspace{-2mm}
			Determine the index of InP, $i_m^{x_2}$, and the index of server in the related InP, $s_m^{x_2}$.\label{Alg_InP_index}\\
			\vspace{-2mm}
			$XP_m^{x_2} = X_{m-1}, \; RemovedState =\{\}$.  \label{Alg_Input_Start} \\
			\vspace{-2mm}
			\If{$x_2 \neq 0$}
			{
			\vspace{-2mm}
				\For{$x_1 \in XP_m^{x_2}$}
				{
				\vspace{-2mm}
					\For{$j=1:|R|$}
					{
					\vspace{-2mm}
						\If{$ (\text{SR}_{m-1}^{x_1}[i_m^{x_2},s_m^{x_2}, j] -r_{u_m,j}^{\rho_{k_m}} < 0)$}
						{
							\vspace{-1mm}
							$XP_m^{x_2} = XP_m^{x_2} -\{x_1\}$.
							\textbf{break}.
							\vspace{-2mm}
						}
						\vspace{-4mm}
					}
					\vspace{-4mm}
				}
				\vspace{-3mm}
				\If{$m \bmod 2 = 0$ \&\& $x_2 \in XP_m^{x_2}$}
				{
					\vspace{-2mm}
					$XP_m^{x_2} = XP_m^{x_2} -\{x_2\}$. \\
					\vspace{-2mm}
				}
				\vspace{-4mm}
			}
			\label{Alg_Input_End}
			\vspace{-2mm}
			\uIf{($XP_m^{x_2} \neq \phi$)}
			{
				\vspace{-2mm}
				\For{$x_1 \in XP_m^{x_2}$}
				{
					\vspace{-2mm}
					compute $\Theta_{m-1,m}^{x_1,x_2}$ using \eqref{DecisionMetric}. \vspace{-1mm}\\
					\vspace{-1mm}
				}
				\vspace{-2mm}
				Compute ${SP}_{m}^{x_2}$ using (\ref{SP_Selection}) and $SR_m^{x_2} = SR_{m-1}^{SP_m^{x_2}}$. \label{Alg_SurvivedPath} \\
				\vspace{-1mm}
				Compute $\Lambda_{m,x_2}^W$ and $\Lambda_{m,x_2}^I$ using \eqref{InP_Server_index}. \label{Alg_Update_Start}\\
				\vspace{-1mm}
				Compute $\phi_{m}^{x_2}$ and $\tau_{m}^{x_2}$ using \eqref{SC_Update} and \eqref{SR_Update}.\\
				\vspace{-1mm}
				\If{$x_2 \neq 0$}
				{
				\vspace{-2mm}
					\For{$j=1:|R|$}
					{
					\vspace{-1mm}
						\scriptsize{$SR_m^{x_2}[i_m^{x_2},s_m^{x_2}, j] = SR_m^{x_2}[i_m^{x_2},s_m^{x_2}, j] - r_{u_m,j}^{\rho_{k_m}}$}.\\
					\vspace{-2mm}
					}
					\vspace{-4mm}
				}
				\vspace{-3mm}
				\label{Alg_Update_End}
			}
			\Else
			{
			\vspace{-2mm}
				Add the  $x_2$ to the $RemovedState$. \label{Alg_RemovedState} \vspace{-2mm}
			}
			\vspace{-4mm}
		}
		\label{Alg_InnerLopp_End}
		\vspace{-3mm}
		\uIf{($RemovedState = X_m$)}
		{\label{Alg_RemovedCheck}
		\vspace{-2mm}
			$\Delta = 0$, \textbf{return}. \label{AlgorithmEnd}
			\vspace{-2mm}
		}
		\vspace{-1mm}
		\Else
		{
			\vspace{-2mm}
			Remove all states in the $RemovedState$ from the $X_m$.\\
			\vspace{-2mm}
		}
		\vspace{-4mm}
	}
	\vspace{-3mm}
	Compute $\xi_{p}^{l,k}, e^{l,k}, N^s_{i,j}[l,k]$ by running Algorithm \ref{Alg2} \label{OutputCompute}.
	\vspace{-1mm}
\end{algorithm}

\begin{algorithm}[!htbp]
	
	\small
		
	\caption{\small{Viterbi Output Computation (VOP) Algorithm}}
	
	\label{Alg2}
	\vspace{-2mm}
	Determine $\zeta$, $\mathcal{P}^I_{L_V}$ and $\mathcal{P}^W_{L_V}$, using (\ref{BP_Calculation}), $a^l = 0$ for $l = 1,2,\ldots, L$.\label{Alg_Survived2}\\

	\vspace{-2mm}
	\For{$m=1:L_V$}
	{
	\vspace{-2mm}
		Determine $k_m$, $u_m$ and $i_{u_m} = \mathcal{P}^I_{L_V}[m], s_{u_m} = \mathcal{P}^W_{L_V}[m]$.\\
		\vspace{-2mm}
		\If{($u_m = 1$)}
		{
			\vspace{-2mm}
			$a^{\rho_{k_m}} = a^{\rho_{k_m}} + 1$. \\
			\vspace{-2mm}
		}
		\vspace{-3mm}
		\For{$j=1:|R|$}
		{
		\vspace{-2mm}
		\label{UsageResource1}
			$N_{i_{u_m}, j}^{s_{u_m}}[\rho_{k_m}, a^{\rho_{k_m}}] = N_{i_{u_m}, j}^{s_{u_m}}[\rho_{k_m}, a^{\rho_{k_m}}] + r_{u_m,j}^{\rho_{k_m}}$.\vspace{-1mm}\\
		}
		\vspace{-1mm}
		\label{UsageResource2}
		\vspace{-1mm}
		$\xi_p^{\rho_{k_m}, a^{\rho_{k_m}}} = \xi_p^{\rho_{k_m}, a^{\rho_{k_m}}}+ \sum_{j=1}^{|R|} r^{\rho_{k_m}}_{u_m,j}\times C_{i_{u_m},j} + DC_{i_{u_m},t_{u_m}^{\rho_{k_m}}}$. \label{PlacementCost1}\\
		\vspace{-1mm}
		\If{($u_m \geq 2$ \&\& $m \bmod 2 = 1$)}
		{
		\vspace{-1mm}
			$\xi_p^{\rho_{k_m}, a^{\rho_{k_m}}} = \xi_p^{\rho_{k_m}, a^{\rho_{k_m}}}+ b^{\rho_{k_m}} \times \Big(C_{i_{u_{m-2}},i_{u_m}}^{s_{u_{m-2}},s_{u_m}} + C_{i_{u_{m-1}},i_{u_m}}^{s_{u_{m-1}},s_{u_m}}\Big)$.\\
			\vspace{-2mm}
		}
		\vspace{-2mm}
		\If{($u_m \geq 2$ \&\& $m \bmod 2 = 0$)}
		{	
		\vspace{-1mm}
			$\xi_p^{\rho_{k_m}, a^{\rho_{k_m}}} = \xi_p^{\rho_{k_m}, a^{\rho_{k_m}}}+ b^{\rho_{k_m}} \times \Big(C_{i_{u_{m-2}},i_{u_m}}^{s_{u_{m-2}},s_{u_m}} + C_{i_{u_{m-3}},i_{u_m}}^{s_{u_{m-3}},s_{u_m}}\Big)$.\\
			\vspace{-2mm}
		}
		\vspace{-2mm}
		\label{PlacementCost2}
		\If{($m \bmod 2 = 0$)}
		{
		\vspace{-1mm}
		\label{FailureProb1}
			$p^{\rho_{k_m}, a^{\rho_{k_m}}} = p^{\rho_{k_m}, a^{\rho_{k_m}}} \times (1-v_{i_{u_m}}v_{i_{u_{m-1}}})$
			\vspace{-2mm}
		}
		\vspace{-2mm}
		\If{($m = U^{\rho_{k_m}}$)}
		{
			\vspace{-1mm}
			$e^{\rho_{k_m}, a^{\rho_{k_m}}} = 1 - p^{\rho_{k_m}, a^{\rho_{k_m}}}$.
			\vspace{-2mm}
		}
		\vspace{-3mm}
		\label{FailureProb2}
	}
	\vspace{-1mm}
\end{algorithm}
\vspace{-6mm}
Algorithm \ref{Alg1} shows the details of the proposed VRSSP algorithm. At the beginning, we initialize the outputs and variables used in the algorithm. The variable $SR_0^0[i,s,j]$ indicates the idle resources of the servers, at the beginning. Then, we have a loop with a length of $L_V$. In each iteration of this loop, we determine the candidate for the main or backup server of a specific VNF. At the beginning of each iteration, in Line \ref{Alg_Service}, we obtain the index of considered service and the index of considered VNF in the corresponding service. Also, the set of states (i.e., candidates) for hosting the corresponding VNF is determined in Line \ref{Alg_State}. In the following, we have an inner loop with the length $|X_m|$ in Lines \ref{Alg_InnerLopp_Start}-\ref{Alg_InnerLopp_End}. In each iteration of the inner loop, we consider one of the placement candidates. In Line \ref{Alg_InP_index}, the index of server and InP for the considered placement candidate, $x_2$ is determined. Then, the set of possible input paths, $XP_m^{x_2}$ is determined in Lines \ref{Alg_Input_Start}-\ref{Alg_Input_End}. For this purpose, we should consider the idle resources of servers of the input paths, $SR_{m-1}^{x_2}$, in the $(m-1)^\text{th}$ stage. If the set of possible input paths to the $(x_2)^\text{th}$ candidate is null, the corresponding server did not have enough resource for hosting the considered VNF and added to a set named $RemovedState$, in Line \ref{Alg_RemovedState}. Otherwise, between all possible input paths to the $(x_2)^\text{th}$ candidate, the path with minimum cost named survived path is selected in Line \ref{Alg_SurvivedPath}. Then, the value of Viterbi related parameters are updated in Lines \ref{Alg_Update_Start}-\ref{Alg_Update_End}.

When all candidates are considered if the $RemovedState$ and $X_m$ are equal (Line \ref{Alg_RemovedCheck}), none of the servers has enough resource and as a result, the rest of services will not be admitted. More precisely, the services $\{k_m, k_m+1, \ldots, K\}$ cannot be admitted because of the insufficient resource of the servers. In such a scenario, the validation metric is set to $0$, and the VRSSP algorithm is terminated, as indicated in Line \ref{AlgorithmEnd}. This scenario can happen only in the odd stages in which we consider main server allocation. This is because that in the even stages, we have no server assignment as an option, and as a result, $XP_m^{x_2}$ could not be null and $X_m \neq RemovedState$. Finally, if the validation metric is $1$, the output of the Algorithm \ref{Alg1} is computed using Algorithm \ref{Alg2} as indicated in Line \ref{OutputCompute}. In Algorithm \ref{Alg2}, we compute the failure probability, $e^{l,k}$, placement cost, $\xi_p^{l,k}$, and usage resources, $N^s_{i,j}[l,k]$, for all services. At the beginning of this algorithm, the Viterbi path is computed in Line \ref{Alg_Survived2}. The usage resources of each service are computed in Lines \ref{UsageResource1}-\ref{UsageResource2}, the placement cost for each service is computed in Lines \ref{PlacementCost1}-\ref{PlacementCost2}, and the failure probability of placement for each service is computed in Lines \ref{FailureProb1}-\ref{FailureProb2}.

{\color{black}The computational complexities of Algorithms 1 and 2 can be computed for a specific input, $V_I = ({A}, \mathcal{\rho}, \mathcal{O})$. The number of stages in the VRSSP algorithm is $L_V$. Also, the total number of the servers of all InPs is $|S| = \sum_{i=1}^{|I|}{|S_i|}$. According to these variables, the computational complexities of Algorithms 1 and 2 are $O(L_V \times |S|^2)$ and $O(L_V)$, respectively. Both of these values grow linearly with increasing the number of the services.}
{\color{black}The memory resources needed for Algorithm \ref{Alg1} is mostly used for the storage of the survived path information which is related to the number of the stages and the number of the states. The resource requirement of Algorithm \ref{Alg1} for input $V_I$ is, $L_V \times |S|$, which grows linearly with the increasing the number of the services. Also, the memory resource requirement of Algorithm \ref{Alg2} is negligible compared to Algorithm \ref{Alg1}.
}

\vspace{-4mm}
\section{Optimal~Policy~Computing}
\label{OptimalPolicyComputing_Section}
In an MDP, the agent will act in such a way as to maximize the long-run reward received. The most straightforward framework is the infinite-horizon discounted model, in which we sum the rewards over the infinite lifetime, but discount them geometrically using discount factor $0 < \gamma < 1$. The agent should act so as to optimize $E \big [ \sum_{n=0}^{\infty}{\gamma^n R_n} \big]$, where $R_n$ is the reward gained in the $n^{\text{th}}$ step. In this model, rewards received earlier in its lifetime have more value to the agent. Even though the infinite lifetime is considered, but the discount factor ensures that the sum is finite. In the infinite-horizon discounted case, we write $V_{\pi}({S})$ for the expected discounted sum of future reward for starting in state $S$ and executing policy $\pi$, which can be computed as
\vspace{-2mm}
\begin{align}
\textstyle V_{\pi}({S}) = R\big({S}, \pi({S})\big) + \gamma \sum_{S^\prime}{P\big({S}, \pi({S}), {S}^\prime\big) V_{\pi}({S}^\prime)} . \label{ValueFunction}
\end{align}

This value function for each policy $\pi \in \Omega_{\pi}$ is the unique simultaneous solution of this set of linear equations, i.e., one equation for each state ${S} \in \Omega_S$. In the infinite-horizon discounted model, for any initial state ${S} \in \Omega_S$, we want to obtain and execute the policy $\pi$ that maximizes value function $V_{\pi}({S})$. We know that there exists a stationary policy, $\pi^*$, that is optimal for every starting state \cite{kaelbling1998planning}. The value function of this policy, $V^*$, can be computed by the set of equations
\vspace{-10mm}
\begin{align}
\hspace{-1mm} \textstyle V^*({S}) = \max_{{A} \in \Omega_A} \Big\{R\big({S}, {A}\big) + \gamma \sum_{S^\prime}{P\big({S}, {A}, {S}^\prime\big) V^*({S}^\prime)} \Big\}.  \label{OptimalValueFunction}
\end{align}

According to \eqref{OptimalValueFunction}, the optimal policy, $\pi^*$, is just a greedy policy with respect to $V^*$. There are many methods for finding optimal policies for MDPs. Here, we would like to explore the value iteration algorithm to find the optimal policy \cite{kaelbling1998planning}. For this purpose, we should first introduce the proposed method for estimating the state of idle resources of the InPs, $\{(\omega_{i,j}^s)\}$, according to the state of active services in $S = \big\{(\lambda^1,\lambda^2,\ldots,\lambda^L),(\sigma^1,\sigma^2,\ldots,\sigma^L)\big\}$.
\vspace{-6mm}
\subsection{State~Estimation~for~InP~Resources}
\label{StateEstimation_SubSection}
As mentioned in Section \ref{State_Set_Part}, we would like to introduce a method for estimating the state of the idle resources of InPs according to the state of active services. Let $\omega_{i,j}^{s,\eta_S}$ denote the idle resource of $j^{\text{th}}$ resource type in $s^{\text{th}}$ server of $i^{\text{th}}$ InP for  $S = \big\{(\lambda^1,\lambda^2,\ldots,\lambda^L),(\sigma^1,\sigma^2,\ldots,\sigma^L)\big\}$, and $\eta_S = 1+\sum_{l=1}^{L}{\sigma^l \times \delta_{\sigma}^l}$, where $\delta_{\sigma}^l$ is defined in \eqref{incomeserviceind}. We know that the state of idle resources is depend on the state of active services and independent of the state of incoming services. We assume that the NO is in the state $S$, takes action $A = \{a^1,a^2,\ldots,a^L\}$ and ends in state $S^{\prime}$. The estimation for the state of the idle resources in state $S^{\prime}$ can be updated as
\vspace{-3mm}
\begin{align}
\omega_{i,j}^{s,\eta_{S^{\prime}}} &= \alpha^{\eta_{S^{\prime}}}  \times \big(\omega_{i,j}^{s,\eta_{S}} - N^s_{i,j}(S,A)\big) + (1-\alpha^{\eta_{S^{\prime}}}) \times \omega_{i,j}^{s,\eta_{S^{\prime}}}, \quad \alpha^{\eta_{S^{\prime}}} = \alpha^{\eta_{S^{\prime}}} \times \mathcal{D}^{\eta_{S^{\prime}}} \label{EtsimFormula},
\end{align}
where $\omega_{i,j}^{s,\eta_{S^{\prime}}}$ denotes the state of the idle resources in state $S^{\prime}$ and $N^s_{i,j}(S,A)$ is the amount of usage resources of $j^{\text{th}}$ resource type in the $s^{\text{th}}$ server of $i^{\text{th}}$ InP for taking action $A$ in state $S$. Finally, $\alpha^{\eta_{S^{\prime}}}$ is the updating factor for estimating the state of the idle resources in state $S^{\prime}$. The value of $\alpha^{\eta_{S^{\prime}}}$ is reduced in each resource estimation of state $S^{\prime}$ using the discount factor $\mathcal{D}^{\eta_{S^{\prime}}}$. The value of $\eta_{S^{\prime}}$ and $N^s_{i,j}(S,A)$ can be computed as
\vspace{-3mm}
\begin{align}
N^s_{i,j}(S,A) &= \textstyle \sum_{l=1}^{L}{\sum_{k=1}^{a^l}{I(F^l-e^{l,k}) \times N^s_{i,j}[l,k]}}, \label{ResourceUsageAll}\\
\eta_{S^\prime} &= 1+\textstyle \sum_{l=1}^{L}{\big(\sigma^l + a^l_r \big) \times \delta_{\sigma}^l}, \; \; a^l_r = \sum_{k=1}^{a^l}{I(F^l-e^{l,k})}, \; A_r = (a^1_r, a^2_r, \ldots, a^L_r) \label{ActionRevise},
\end{align}
where $N^s_{i,j}[l,k]$ and $e^{l,k}$ are the outputs of running VRSSP algorithm with input of $V_I = (A, \rho, \mathcal{O})$ in which $\mathcal{O} = \{(\omega_{i,j}^{s,\eta_{S}})\}$ and $\rho = (\rho_1, \rho_2, \ldots, \rho_{K})$ is an arbitrary placement arrangement of the services which should be admitted according to action $A$. Also, $a_r^l$ indicates the number of service of $l^{\text{th}}$ type which can be admitted according to the output of the VRSSP algorithm.

According to the introduced updating formula for the state of the idle resources in \eqref{EtsimFormula}, we should determine the initial value of $\omega_{i,j}^{s,\eta_{S}}$, which is set to zero. On the other hand, for state $S = \big\{(\lambda^1,\lambda^2,\ldots,\lambda^L),(0,0,\ldots,0)\big\}$ which means that there is no active service in the system and we have $\omega_{i,j}^{s,\eta_{S}} = 0$. In such a state, with taking different feasible action $A \in \Omega_A$ and ends in different state $S^{\prime} \in \Omega_S$, the value of $\omega_{i,j}^{s,\eta_{S^{\prime}}}$ can updated using \eqref{EtsimFormula}.
\vspace{-6mm}
\subsection{Placement~Arrangement~for~Actions}
\label{Arrangment_SubSection}
As mentioned in Section \ref{StaticAlgorithm_Section}, one of the inputs of the VRSSP algorithm is the placement arrangement of the services which should be admitted according to the taken action $A$. The placement arrangement for action $A = \{a^1,a^2, \ldots, a^l\}$ with $K = \sum_{l=1}^{L}{a^l}$ is indicated as $\rho = (\rho_1, \rho_2, \ldots, \rho_{K})$, where $\rho_k = l$, means that the type of $k^{\text{th}}$ service for placement is $l$. We want to find optimum the optimum placement arrangement for each feasible action $A$ in each state $S$. For this purpose, we define $\rho^{S,A}$ and $\rho_{\text{opt}}^{S,A}$ as the set of possible placement arrangement and the optimum placement arrangement of taking action $A$ in state $S$.
The possible placement arrangements for each feasible action $A$ in each state $S$ are evaluated within the value iteration algorithm and the optimum placement arrangement is determined. In this way, $R_{\text{opt}}^{S,A}$ is the reward of optimum placement arrangement for taking action $A$ in state $S$. The values of $R_{\text{opt}}^{S,A}$ and $\rho_{\text{opt}}^{S,A}$ are updated, when $\rho \in \rho^{S,A}$ which is one of the possible placement arrangements for taking action $A$ in state $S$, is considered within the value iteration algorithm. More precisely, if the value of $R(S,A)$ is greater than the value of $R_{\text{opt}}^{S,A}$, the values of $\rho_{\text{opt}}^{S,A}$ and $R_{\text{opt}}^{S,A}$ are updated. The value of $R(S,A)$ is determined using \eqref{RewardFunction} where the values of $e^{l,k}$ and $\xi_{p}^{l,k}$ are the output of running VRSSP algorithm with input of  $V_I = (A, \rho, \mathcal{O})$ in which $\mathcal{O} = \{(\omega_{i,j}^{s,\eta_{S}})\}$.

\begin{algorithm}[!htbp]
	
	\begingroup
	\small
	\caption{\small{VRSSP-based Value Iteration (VVI) Algorithm}}
	\label{Alg_VVI}
	\vspace{-1mm}
	\textbf{Service input:} $\Upsilon^l = \big\{F^l, d^l, b^l, U^l, f^l, \lambda^l_{\text{max}}, r^l_{u,j}, t^l_u \big\}, \; L$.\label{AlgServInput} \\
	\vspace{-2mm}
	\textbf{MDP related parameter:} $q^l, \; \sigma^l_{\text{max}}, \; \gamma, \; \epsilon, \; |O|$. \label{AlgMDPInput}\\
	\vspace{-2mm}
	Compute the state set $\Omega_S = \Omega_S^D \times \Omega_S^I$ using \ref{ExistState}-\ref{IncomingeState} and state transition matrix, $\textbf{P}$, using \ref{Transition_Matrix_Partial1}-\ref{Transition_Matrix_Partial4}.\label{AlgTransition}\\
	\vspace{-2mm}
	$ \omega_{i,j}^{s,\eta_S} = 0, \; \; \alpha^{\eta_S} = 1, \; \; \mathcal{D}^{\eta_S} = 0.5$ and $V_0(S) = 0 \;\; \forall S \in \Omega_S, \;\; n = 1$. \label{MDPInit}\\
	\vspace{-2mm}
	\For{$S \in \Omega_S$}
	{
	\vspace{-2mm}
	\label{AlgArragementGeneStart}
		\small{Compute the set of feasible action, $\Omega_A^f$, using \ref{FeasibleAction}}.\\
		\vspace{-2mm}
		\For{$A \in \Omega_A^f$}
		{
			\vspace{-1mm}
			$R_{\text{opt}}^{S,A} = 0, \; \rho^{S,A} = \{\}, \; K = \sum_{l=1}^{L}{a^l}, \; STV =\{\}$.\\
			\vspace{-2mm}
			\For{($l =1:L$)}
			{
			\vspace{-2mm}
				\For{($a=1:a^l$)}
				{
					\vspace{-1mm}
					Add $l$ to the $STV$. \vspace{-3mm}\\
				}
				\vspace{-3mm}
			}
			\vspace{-3mm}
			\For{($o =1:|O|$)}
			{
				\vspace{-2mm}
				$Perm =$ A Random permutation of $STV$ and add $Perm$ to the $\rho^{S,A}$.\vspace{-2mm}\\
			}
			\vspace{-4mm}			
		}
		\vspace{-3mm}
	}
	\label{AlgArragementGeneEnd}
	\vspace{-3mm}
	\While {$\big(n \geq 2$ \&\& $\big|V_n(S) - V_{n-1}(S)\big| \geq \epsilon\big)$}
	{
	\vspace{-2mm}
	\label{ALgValueIterationStart}
		\For{$S=(\lambda,\sigma) \in \Omega_S$}
		{
			\vspace{-2mm}
			Compute the set of feasible action, $\Omega_A^f$, using \ref{FeasibleAction}. \label{AlgFeasibleAct}\\
			\vspace{-1mm}
			$\eta_S = 1+\sum_{l=1}^{L}{\sigma^l \times \delta_{\sigma}^l}, \; \omega_{i,j}^s = \omega_{i,j}^{s,\eta_S} \;\; \forall i, j, s$ and $A_r = (0,0,\ldots,0)$.\label{AlgRemainRescall}\\
			\vspace{-1mm}
			\For{$A \in \Omega_A^f$}
			{
				\vspace{-2mm}
				\label{AlgQualityStart}
				Select $\rho \in \rho^{S,A}, \rho^{S,A} = \rho^{S,A} - \{\rho\}$ and $R(S,A) = 0$.\label{AlgArrangementSelect}\\
				\vspace{-2mm}
				$V_I = ({A}, \rho, \omega_{i,j}^s)$ and $ (\Delta, \xi_{p}^{l,k}, e^{l,k}, N^s_{i,j}[l,k]) = $ Algorithm\ref{Alg1}$(V_I)$.\label{AlgVRSSPCall}\\
				\vspace{-2mm}
				\If{($\Delta = 1$)}
				{
					\vspace{-2mm}
					Compute $R(S,A)$ using \eqref{RewardFunction}.\label{AlgRewardCalcul}\\
					\vspace{-2mm}
					Compute $\eta_{S^{\prime}}$ and $A_r$ using \eqref{ActionRevise}.\label{AlgRevisedAction}\\
					\vspace{-2mm}
					Update $\omega_{i,j}^{s,\eta_{S^{\prime}}}$ and $\alpha^{\eta_{S^{\prime}}}$ using \eqref{ResourceUsageAll} and \eqref{EtsimFormula}.\label{AlgResUpdated}\\
					\vspace{-1mm}
					\If{($R_{\text{opt}}^{S,A} \leq R(S,A)$)}
					{
						\vspace{-1mm}
					\label{AlgOrderUpdateStart}
						$R_{\text{opt}}^{S,A} = R(S,A), \; \rho_{\text{opt}}^{S,A} = \rho$. \\ \vspace{-3mm}
					}
					\vspace{-4mm}
					\label{AlgOrderUpdateEnd}
				}
				\vspace{-3mm}
				$Q_n^A(S) = R(S,A)+ \gamma \times \sum_{S^{\prime} \in \Omega_S}{P(S,A_r,S^{\prime}) \times V_{n-1}(S^{\prime}})$.\vspace{-4mm} \label{AlgQualityCompute}\\
			}
			\label{AlgQualityEnd}
			\vspace{-2mm}
			$A = \argmax_{A \in \Omega_A^f}{Q_n^A(S)}$.\label{AlgValueUpdate1}\\
			\vspace{-1mm}
			$V_n(S) = Q_n^A(S), \; O_n(S) = O_{\text{opt}}^{S,A}$.\label{AlgValueUpdate2} \vspace{-4mm}
		}
		\vspace{-4mm}
	}
	\label{ALgValueIterationEnd}
	\endgroup
	\vspace{-2mm}
\end{algorithm}

In Algorithm \ref{Alg_VVI} the VRSSP-based Value Iteration (VVI) algorithm for finding the optimal policy of MDP is indicated. In Lines \ref{AlgServInput} and \ref{AlgMDPInput}, the characteristics of service types and MDP-related parameters are determined. Then, the set of state space and transition matrix are computed in Line \ref{AlgTransition}. The state of idle resources in the InPs and the value function of each state are initialized in Line \ref{MDPInit}. The set of possible placement arrangement for each feasible action in each state, $\rho^{S,A}$ is computed in Lines \ref{AlgArragementGeneStart}-\ref{AlgArragementGeneEnd}. For this purpose, the $STV$ is a vector which indicates the set of service types which should be admitted according to action $A$. For example, if $A = (1, 2, 0, 1)$, we have $STV = \{1, 2, 2, 4\}$ which means that one service of first service type, two services of second service type and one service of forth service type should be admitted. According to $STV$, there are 12 possible placement arrangements for the given action $A$. On the other hand, because of the structure of the VRSSP algorithm, each one of these placement arrangements can lead to a different output which changes the performance of the proposed algorithm. As a result, we can obtain the optimal placement arrangement within the value iteration process.

In Lines \ref{ALgValueIterationStart}-\ref{ALgValueIterationEnd}, the value iteration is executed for finding the optimal action. In each iteration of the value iteration algorithm, the value function of each $S \in \Omega_S$ is updated. In this way, the set of feasible actions for state $S$ is determined in Line \ref{AlgFeasibleAct} and the state of idle resources in the InPs is obtained in Line \ref{AlgRemainRescall}. Then, the value of $Q_n^A(S)$ which is the $n^{\text{th}}$ step value of starting in state $S$, taking action $A$, then continuing with the optimal $(n-1)^{\text{th}}$ step nonstationary policy, is determined in Lines \ref{AlgQualityStart}-\ref{AlgQualityEnd}. For this purpose, one placement arrangement is selected in Line \ref{AlgArrangementSelect}, the input of VRSSP algorithm is determined, and VRSSP algorithm runs in Line \ref{AlgVRSSPCall}. In case of valid input ($\Delta = 1$), the reward of taking action $A$ in state $S$ is computed in Line \ref{AlgRewardCalcul}, the number of admitted services according to the output of VRSSP algorithm is computed in Line \ref{AlgRevisedAction} and the index of next state, $\eta_{S^{\prime}}$ and $A_r$ are determined in Line \ref{AlgRevisedAction}. Then, the state of idle resources and the values of $R_{\text{opt}}^{S,A}$ and $O_{\text{opt}}^{S,A}$ are updated in Lines \ref{AlgResUpdated}-\ref{AlgOrderUpdateEnd}. The value of $Q_n^A(S)$ is computed in Line \ref{AlgQualityCompute} and the values of $V_n(S)$ and $O_n(S)$ are updated in Lines \ref{AlgValueUpdate1}-\ref{AlgValueUpdate2}, where $O_n(S)$ is the optimal placement arrangement for the optimal action of $n^{\text{th}}$ step in state $S$.

{\color{black}For evaluating the complexity of Algorithm \ref{Alg_VVI}, we compute the computational complexity of each iteration of the value iteration process. The computational complexity of each iteration is $O(|\Omega_S| \times |\Omega_A| \times |U| \times |S|^2)$, where $|\Omega_S|$ is the size of state set, $|\Omega_A|$ is the size of action set, $|S|$ is the total number of the servers of all InPs, and $|U| = 2 \times \sum_{l=1}^{L}{U^l \times \lambda^l_{\text{max}}}$ is the maximum number of stages in the VRSSP algorithm.}
{\color{black}The memory usage of Algorithm \ref{Alg_VVI} is mostly due to the storage of the placement arrangement information and the value function of the states. The memory resource of the placement arrangement information is, $|\rho^{S,A}| \times |O| \times |A|$, where $|\rho^{S,A}| = |\Omega_S| \times |\Omega_A|$, $|O|$ is the maximum number of the permutations which are evaluated for each action during Algorithm \ref{Alg_VVI}, and $|A| = \sum_{l=1}^{L}{\lambda^l_{\text{max}}}$ is the maximum number of the services which can be admitted according to an action. The memory resource of the value function of the states is, $|\Omega_S| \times |\Omega_A|$, which is related to the required memory for $Q_n^A(S)$. The memory resources of $Q_n^A(S)$ is negligible compared to the memory resource of the placement arrangement information.Therefore, the memory resource of Algorithm \ref{Alg_VVI} is, $|\rho^{S,A}| \times |O| \times |A|$.
}

\begin{figure}[!t]
\centering
\begin{tikzpicture}[scale=0.6, every node/.style={scale=0.6}]

\draw [-, thick] (-2.5,7.2) -- (0.5,7.2);
\draw [-, thick] (-2.5,6.2) -- (0.5,6.2);
\draw [-, thick] (-2.5,5.2) -- (0.5,5.2);
\draw [-, thick] (-2.5,4.2) -- (0.5,4.2);

\draw [-, thick] (0.5,7.2) -- (0.5,6.2);
\draw [-, thick] (0.5,6.2) -- (0.5,5.2);
\draw [-, thick] (0.5,5.2) -- (0.5,4.2);

\node[text width=0.5cm, minimum height=0.5cm] at (-0.2, 6.7) {\Large{$,\lambda_{n}^1$}};
\node[text width=0.5cm, minimum height=0.5cm] at (-1.6, 6.7) {\Large{$,\lambda_{n+1}^1$}};
\node[text width=0.5cm, minimum height=0.5cm] at (-2.4, 6.7) {\Large{$\ldots$}};

\node[minimum width=0.5cm, minimum height=0.5cm] at (-0.5, 5.7) {\Large{$\ldots$}};

\node[text width=0.5cm, minimum height=0.5cm] at (-0.2, 4.7) {\Large{$,\lambda_{n}^L$}};
\node[text width=0.5cm, minimum height=0.5cm] at (-1.6, 4.7) {\Large{$,\lambda_{n+1}^L$}};
\node[text width=0.5cm, minimum height=0.5cm] at (-2.4, 4.7) {\Large{$\ldots$}};

\draw [->, thick] (0.5,5.7) -- (1.5,5.7);
\draw [ultra thick] (2.2,5.7) circle (0.7cm);
\node[minimum width=0.5cm, minimum height=0.5cm] at (2.2, 5.7) {\Huge{+}};

%
%
%

\draw [->, thick] (2.9,5.7) -- (6,5.7);
\node[text width=5cm, minimum height=1cm] at (5.5, 6.2) {\large{$S_n=\{\lambda_n,\sigma_n\}$}};

\draw [ultra thick] (6.7,5.7) circle (0.7cm);
\node[minimum width=0.5cm, minimum height=0.5cm] at (6.7, 5.7) {\Huge{+}};

\draw [->, thick] (7.4,5.7) -- (8.4,5.7);
\node[draw,  thick, fill=blue!20, align = center, text width=3.2cm, text height=0.4cm] at (10.15,5.7) {\large{Determine $A_n$ and $\rho_n$ for $S_n$}};

\draw [->, thick] (11.9,5.7) -- (13,5.7);
\draw [ultra thick] (13.7,5.7) circle (0.7cm);
\node[minimum width=0.5cm, minimum height=0.5cm] at (13.7, 5.7) {\Huge{+}};

\draw [->, thick] (14.4,5.7) -- (15,5.7);
\node[draw,  thick, fill=blue!20, align = center, text width=3.8cm, text height=0.4cm] at (17.05,5.7) {\large{Run VRSSP with $V_I=(A_n,\rho_n,O_n)$}};

\draw [->, thick] (19.1,5.7) -- (19.7,5.7);
\node[draw,  thick, fill=blue!20, align = center, text width=3.8cm, text height=0.4cm] at (21.7,5.7) {\large{Service admission, update $\omega_{i,j}^s, \sigma_n$}};

\draw [dashed,red, thick] (8,6.8) -- (24.1,6.8);
\draw [dashed,red, thick] (8,6.8) -- (8,4.6);
\draw [dashed,red, thick] (8,4.6) -- (24.1,4.6);
\draw [dashed,red, thick] (24.1,6.8) -- (24.1,4.6);
\draw [->, thick] (15.9, 7) -- (18, 7.5);
\node[minimum width=0.5cm, minimum height=0.5cm] at (20.6, 7.5) {\Large{Beginning of $n^{\text{th}}$ slot}};

\draw [-, thick] (24.5,5.68) -- (23.75,5.68);
\draw [-, thick] (24.5,5.7) -- (24.5,2.15);
\draw [->, thick] (24.5,2.15) -- (23.75,2.15);
\node[draw,  thick, fill=blue!20, align = center, text width=3.8cm, text height=0.4cm] at (21.7,2.15) {\large{Determine departure services}};

\draw [->, thick] (19.7,2.15) -- (19.1,2.15);
\node[draw,  thick, fill=blue!20, align = center, text width=3.8cm, text height=0.4cm] at (17.05,2.15) {\large{Update $\omega_{i,j}^s$ and determine $\sigma_{n+1}$}};

\draw [dashed,red, thick] (14.6, 3.2) -- (24.1, 3.2);
\draw [dashed,red, thick] (14.6, 3.2) -- (14.6, 1.05);
\draw [dashed,red, thick] (14.6,1.05) -- (24.1,1.05);
\draw [dashed,red, thick] (24.1,3.2) -- (24.1,1.05);
\draw [->, thick] (14.5, 1.8) -- (12.6, 1.3);
\node[minimum width=0.5cm, minimum height=0.5cm] at (10.7, 1.4) {\Large{End of $n^{\text{th}}$ slot}};

\draw [-, thick] (13.7,2.15) -- (15.,2.15);
\draw [->, thick] (13.7,2.15) -- (13.7,5);
\node[minimum width=0.5cm, minimum height=0.5cm] at (15.3, 3.8) {\Large{$O_n = \{\omega_{i,j}^s\}$}};

\draw [-, thick] (13.7,2.15) -- (2.2,2.15);
\draw [->, thick] (2.2,2.15) -- (2.2,5);
\node[minimum width=0.5cm, minimum height=0.5cm] at (4.5, 3.8) {\Large{$\sigma_n = (\sigma_n^1, \ldots, \sigma_n^L)$}};

\draw [->, thick] (6.7,7.4) -- (6.7,6.4);
\node[draw,  thick, fill=blue!20, align = center, text width=2.5cm, text height=0.4cm] at (6.7,8) {\large{Run VVI algorithm}};
\draw [->, thick] (6.7,9.75) -- (6.7,8.75);

\draw [ultra thick] (6.7,10.45) circle (0.7cm);
\node[minimum width=0.5cm, minimum height=0.5cm] at (6.7, 10.45) {\Huge{+}};

\draw [->, thick] (4,10.45) -- (6,10.45);
\node[draw,  thick, fill=blue!20, align = center, text width=3.5cm, text height=0.4cm] at (2.1,10.45) {\large{Service Types Information}};
\draw [-, thick] (0.2,10.45) -- (-1.2,10.45);
\draw [->, thick] (-1.18,10.45) -- (-1.18,7.2);

\draw [->, thick] (10,10.45) -- (7.4,10.45);
\node[inner sep=0pt] (russell) at (15,10.5)
    {\includegraphics[width=0.6\textwidth]{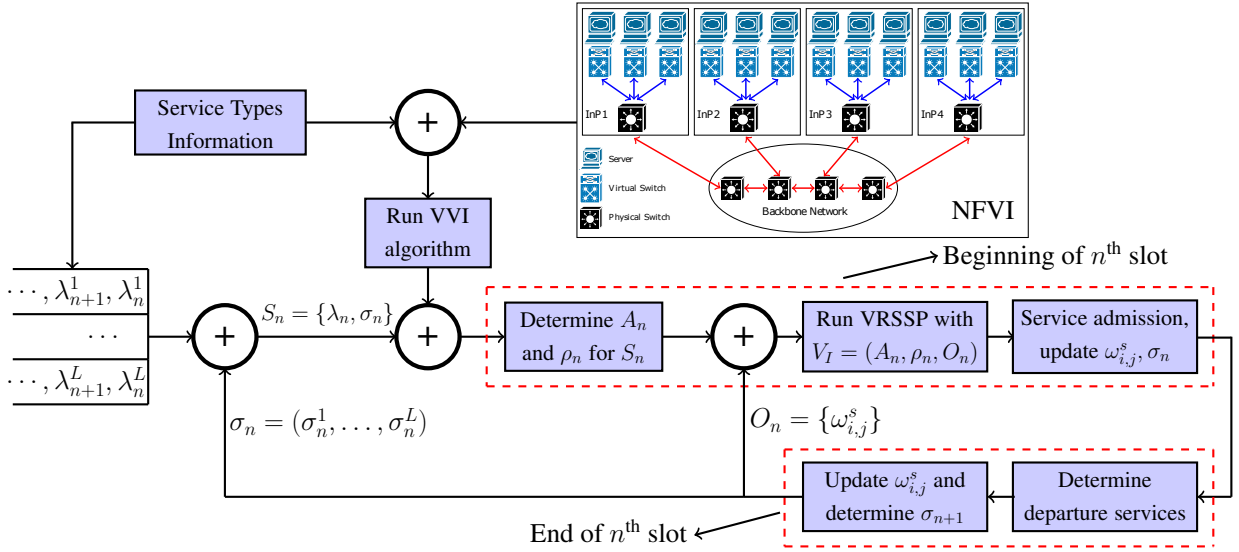}};

\draw (10,7.9) rectangle (20, 13.1);

\node[minimum width=0.5cm, minimum height=0.5cm] at (19, 8.5) {\Large{NFVI}};

\end{tikzpicture}
\vspace{-10mm}
\caption{Dynamic reliability-aware service placement procedure.}
\label{SystemModel_Figure_Tikz}
\vspace{-10mm}
\end{figure}

{\color{black}In Fig. \ref{SystemModel_Figure_Tikz}, the proposed model for dynamic reliability-aware service placement is shown. As seen in this figure, VVI algorithm takes the characteristics of service types and NFVI as the inputs to determine the optimal service admission and placement policy. At the beginning of each slot, NO takes the state of active services and the state of incoming services as the input to determine the optimal action $(A_n, \rho_n)$. Then, according to the state of the NFVI's resources and the optimal action, the VRSSP algorithm determines the admitted services and their placement. Then, the state of the active services and the NFVI's resources are updated. All of these tasks are conducted at the beginning of each slot. At the end of each slot, the terminated services are determined and the state of the active services and the NFVI's resources are updated.}

\vspace{-3mm}
\section{Numerical~Result}
\label{NumericalResult_Section}
In this section, we would like to evaluate the performance of the proposed MDP-based model. We consider three performance metrics, including the placement cost, the number of backup servers, and the admission ratio. The placement cost for each service is introduced in \eqref{Objective_Function}. The number of backup servers indicates the number of additional servers used to meet the required reliability. The admission ratio is defined as the ratio of the number of accepted services with the required reliability to the number of incoming services. We compare the performance of the proposed model with five static methods for reliability-aware service placement, including VRSSP, MinResource, MinReliability, CERA, and RedundantVNF. These algorithms consider the service placement in each slot using the idle resources without concerning about the effects of decisions on the next slots. In the VRSSP algorithm introduced in Algorithm \ref{Alg1}, NO tries to admit all of the arrival services during the $(n-1)^{\text{th}}$ slot, at the beginning of $n^{\text{th}}$ slot. 

In the MinResource, MinReliability, and CERA methods, NO first allocates the main servers to all incoming services. Then, it endeavors to meet the reliability requirement of each service by allocating backup servers. The policy of allocating the main servers in these three methods is minimizing the placement cost, and the main difference between them is on the backup server allocation algorithm. In MinResource method, for each service, NO selects the VNF with the minimum required resource and then chooses a backup server for the given VNF in a way that the required reliability is met. Otherwise, a server with the highest reliability and sufficient resource is allocated as a backup \cite{ding2017enhancing}. In MinReliability method, for each service, NO selects the VNF with the minimum reliability and then chooses a backup server for the given VNF in a way that the required reliability is met. Otherwise, the selected VNF is assigned to a server with the highest reliability and sufficient resource \cite{fan2015grep}. The CERA method employs a metric named cost importance measure for VNF selection for backup and backup placement in each iteration until the required reliability is met \cite{ding2017enhancing}.
{\color{black}The RedundantVNF, introduced in \cite{qu2017reliability}, considers iterative backup adding to the services according to the reliability of each VNF in each service. It is worth noting that this algorithm simultaneously performs main and backup server placement.}

\begin{figure}
	\centering
	\begin{subfigure}[b]{0.495\textwidth}
		\includegraphics[scale=0.24]{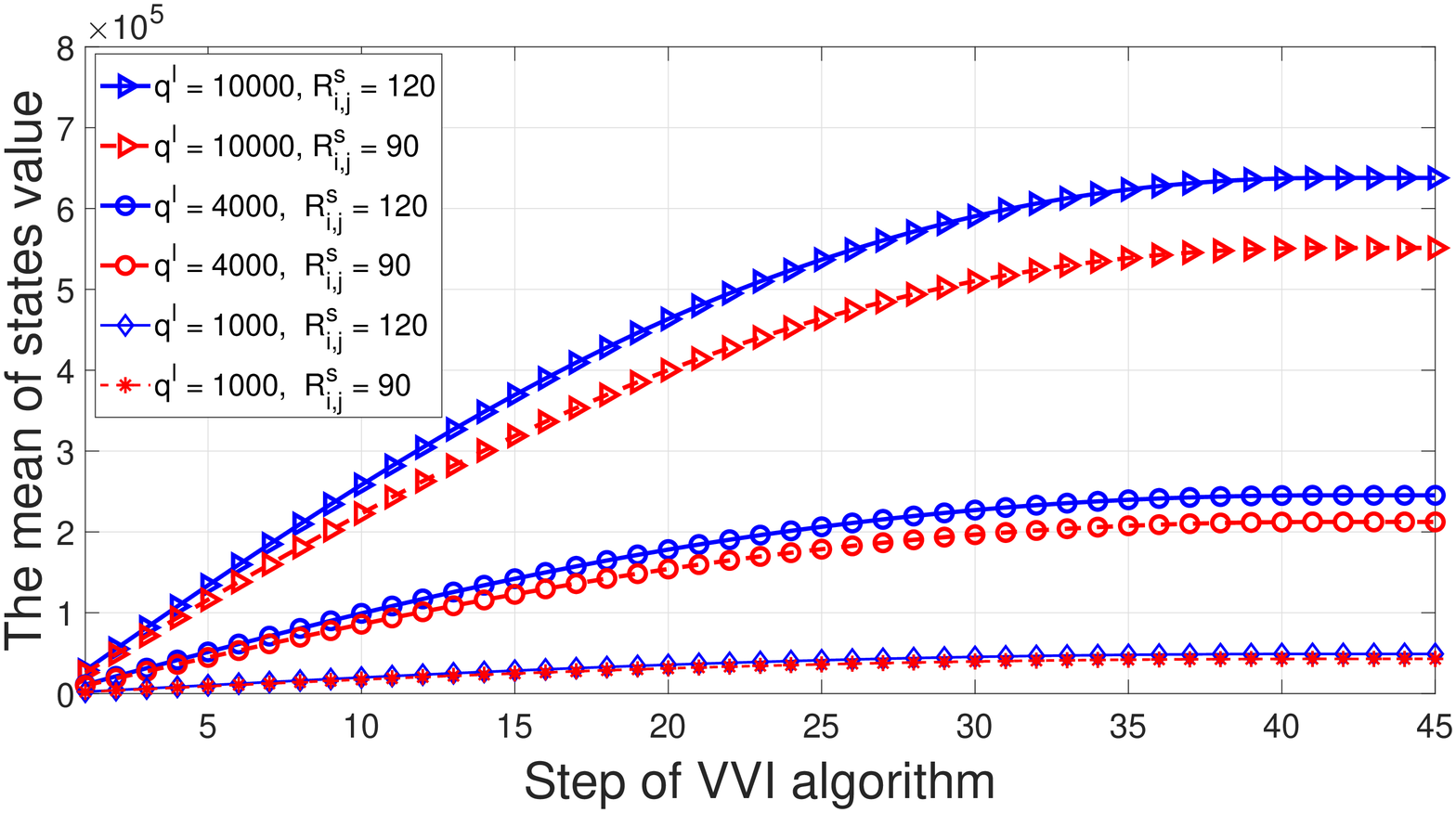}
		\vspace{-.1in}
		\renewcommand{\captionfont}{\small}
		\caption{\scriptsize{Convergence in case of changing service admission reward, $q^l$.}}
		\label{Convergence_ServiceReward}
	\end{subfigure}
	\begin{subfigure}[b]{0.495\textwidth}
		\includegraphics[scale=0.24]{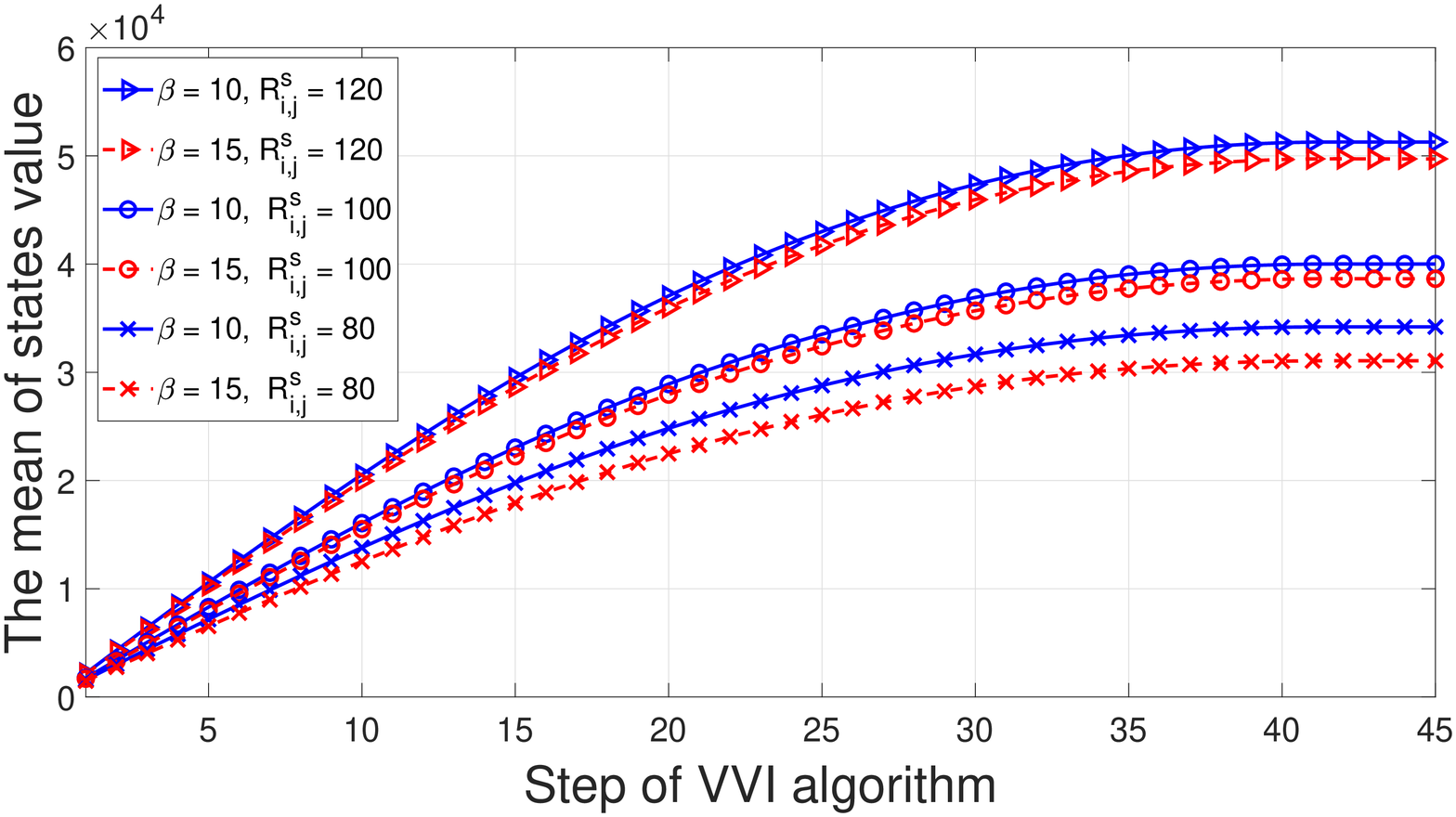}
		\vspace{-.1in}
		\renewcommand{\captionfont}{\small}
		\caption{\scriptsize{Convergence in case of changing server cost parameter, $\beta$.}}
		\label{Convergence_ServerCost}
	\end{subfigure}
	\vspace{-12mm}
	\label{ConvergenceAnalysis}
	\caption{\small{Convergence of the proposed VVI for different scenarios of service admission reward, $q^l$, departure probability $d^l$, and server cost parameter, $\beta$.}}
	\vspace{-10mm}
\end{figure}

\vspace{-4mm}
\subsection{Convergence~of~VVI~Algorithm}
\label{SecSimulation_Convergence}
Now, we investigate the convergence of Algorithm \ref{Alg_VVI} for finding the optimal policy in different scenarios. We first introduce the simulation setup. We consider an NFV-enabled NO which would like to deliver four service types. The reliability requirement of each type is among $\{96, 97, 98, 99\}$, according to the SLA of Google Apps \cite{google_SLA}. The SFC of each service type consists of three to six VNFs, and the resource demand of each VNF is randomly generated between 20 and 30 units. The number of arrival services of different types is a random number between zero and two services in each slot. The NFVI consists of seven InPs with reliability levels from $93\%$ to $99\%$ with $1\%$ step. For each InP, we consider three servers with equal reliability. For evaluating the performance of the MDP model, different values for the resource capacity of each server between 80 and 120 is examined.
{\color{black}It is worth noting that Algorithm \ref{Alg_VVI} is performed offline. Therefore, there is no need for execution of Algorithm \ref{Alg_VVI} during the slot.}

In Fig. \ref{Convergence_ServiceReward}, the convergence of the mean state value, $V^M_n = \sum_{S \in \Omega_S}{V_n(S)}/|\Omega_S|$, for the different values of the service admission reward, $q^l$, and two different values for the capacity of servers, is shown.
In Fig. \ref{Convergence_ServerCost}, the convergence of $V^M_n$, for two different values of parameter $\beta$ used in \eqref{ServerCostvsReliability}, and three different values for the capacity of servers, is indicated.
As seen in these figures, the convergence rate is not affected by changing the value of $q^l$, the value of $\beta$, and the resource capacity of servers.
However, by increasing the value of $q^l$ and the resource capacity of the servers, the value of $V^M_n$ is gained, which is expected.
As seen in Fig. \ref{Convergence_ServerCost}, increasing the value of $\beta$ leads to reduction of $V^M_n$. According to \eqref{ServerCostvsReliability}, increasing the value of $\beta$ leads to increase in server cost, which increases the placement cost and reduces the value of $V^M_n$.

\vspace{-6mm}
\subsection{Performance~Comparison~of~MDP~and~Static~Algorithm}
\begin{figure}[t]
	\centering
	\begin{subfigure}[b]{0.495\textwidth}
		\includegraphics[scale=0.23]{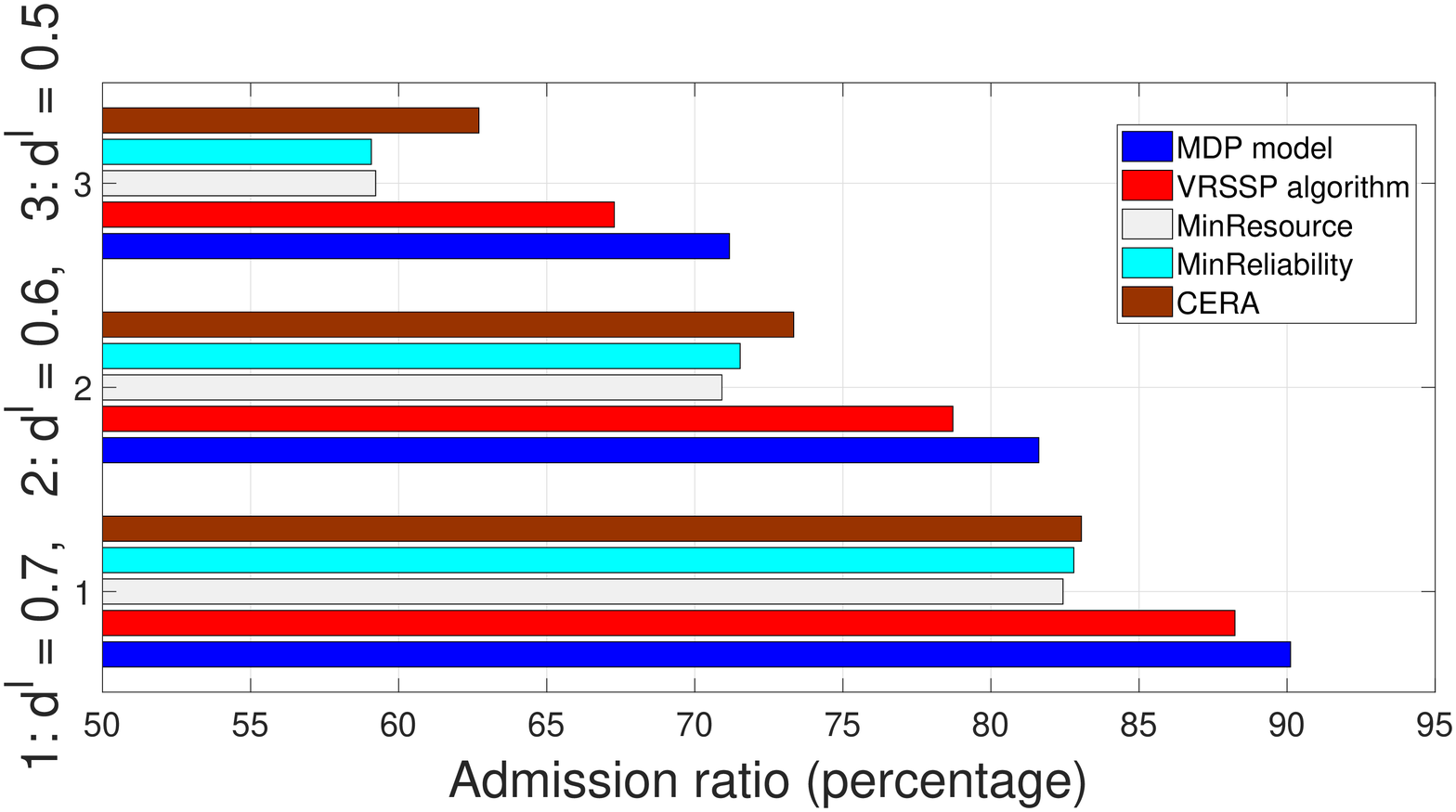}
		\vspace{-.1in}
		\renewcommand{\captionfont}{\small}
		\caption{\scriptsize{Admission ratio for different values of $d^l$.}}
		\label{PerformanceEval_DepartureProb_AdmitRatio}
	\end{subfigure}
	\begin{subfigure}[b]{0.495\textwidth}
		\includegraphics[scale=0.23]{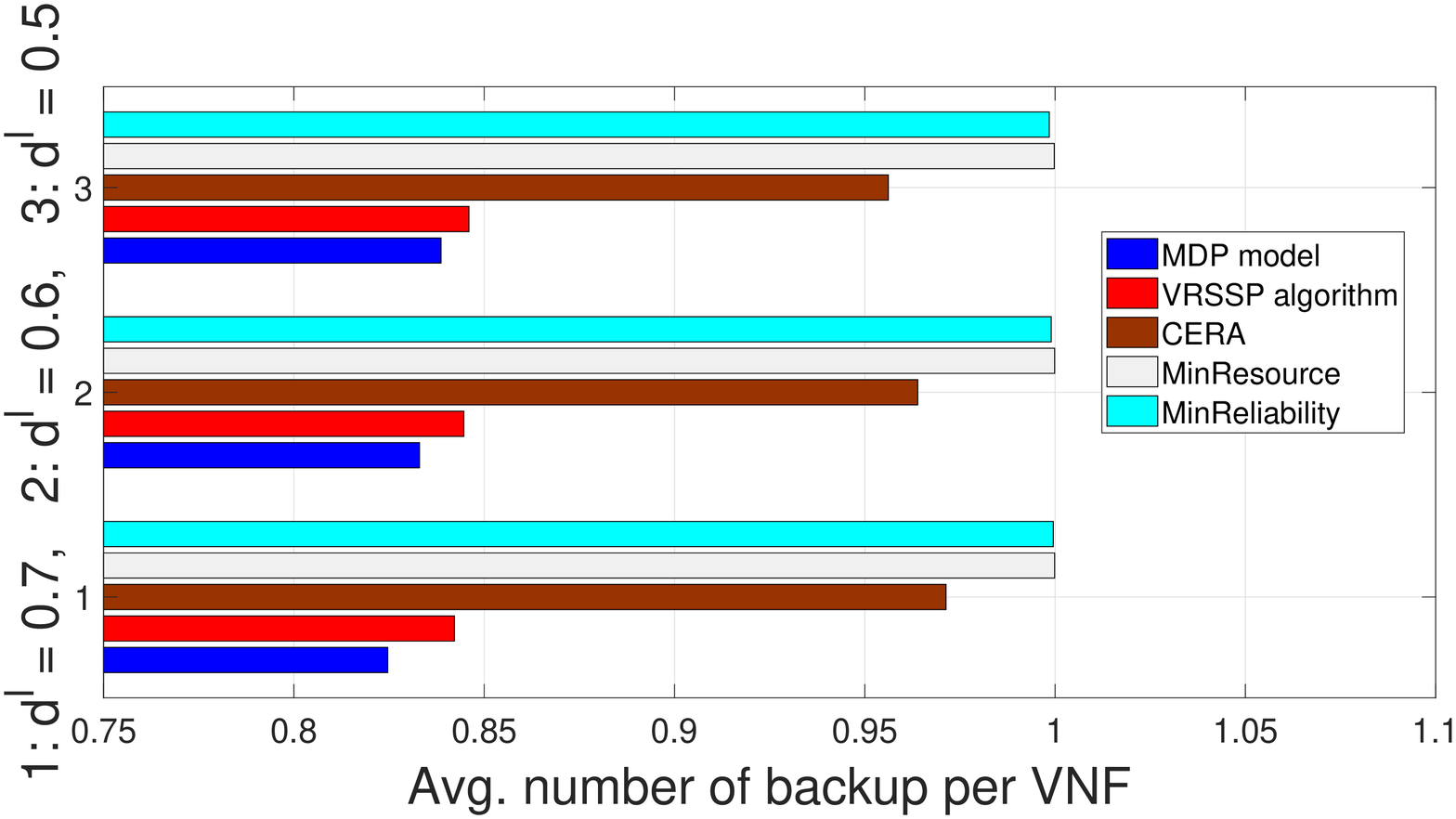}
		\vspace{-.1in}
		\renewcommand{\captionfont}{\small}
		\caption{\scriptsize{Number of backups per VNF for different values of $d^l$.}}
		\label{PerformanceEval_DepartureProb_BackupNum}
	\end{subfigure}
	\vspace{-12mm}
	\caption{\small{Performance of MDP model and static methods for different values of service departure probability, $d^l$.}}		
	\label{PerformanceEval_DepartureProb}
	\vspace{-8mm}
\end{figure}

Now, we would like to evaluate the performance of MDP model by changing the departure probability, $d^l$, and the server resource, $R^s_{i,j}$. The simulation setup is the same as the introduced setup in Section \ref{SecSimulation_Convergence}.
In Figs. \ref{PerformanceEval_DepartureProb_AdmitRatio}-\ref{PerformanceEval_DepartureProb_BackupNum}, the admission ratio and average number of backup servers per VNF for the MDP model and four static methods, by changing the service departure probability are indicated, respectively. The simulation is conducted by assumption of $q^l = 4000, \beta = 15, R^s_{i,j} = 80$.
As seen in Fig. \ref{PerformanceEval_DepartureProb_AdmitRatio}, the admission ratio of MDP model is remarkably higher than the other algorithm, especially by decreasing the value of $d_l$. By decreasing the value of $d^l$, the admitted services last for more number of slots. In such a scenario, the MDP model is much more efficient for improving the admission ratio.
On the other hand, for a large value of $d^l$, the MDP model can be much more efficient for improving the average number of backups, and by increasing the value of $d_l$, the improvement of the MDP model in terms of the average number of the backups is gained, as indicated in Fig. \ref{PerformanceEval_DepartureProb_BackupNum}. This improvement in the average number of backups is due to the fact that for large values of $d^l$, the MDP model can find an appropriate placement for each service, by admitting the services in proper states.

\begin{figure}
	\centering
	\begin{subfigure}[b]{0.495\textwidth}
		\includegraphics[scale=0.25]{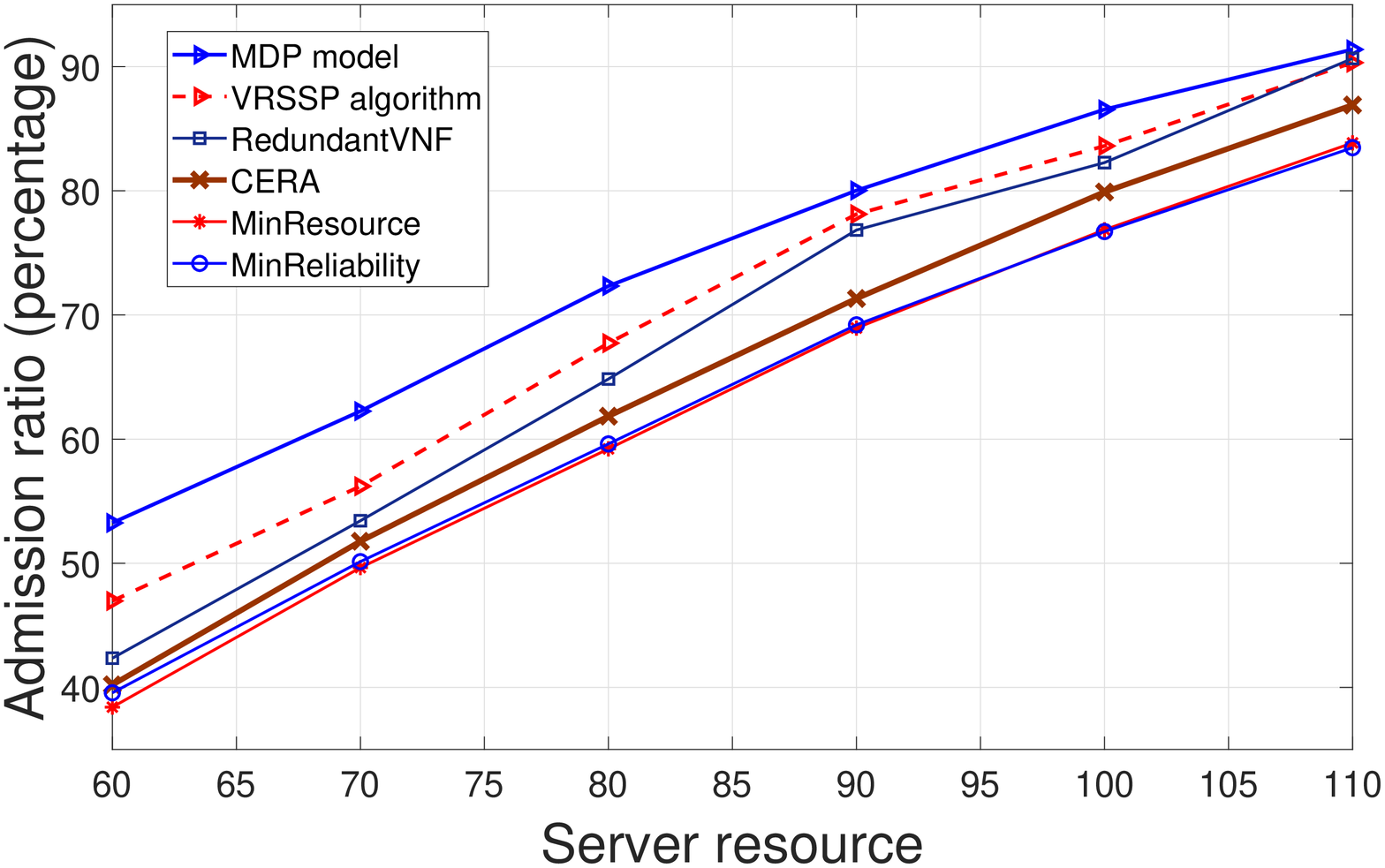}
		\vspace{-.1in}
		\renewcommand{\captionfont}{\small}
		\caption{\scriptsize{Admission ratio for different values of server resource.}}
		\label{PerformanceEval_ServerResource_AdmitRatio}
	\end{subfigure}
	\begin{subfigure}[b]{0.495\textwidth}
		\includegraphics[scale=0.25]{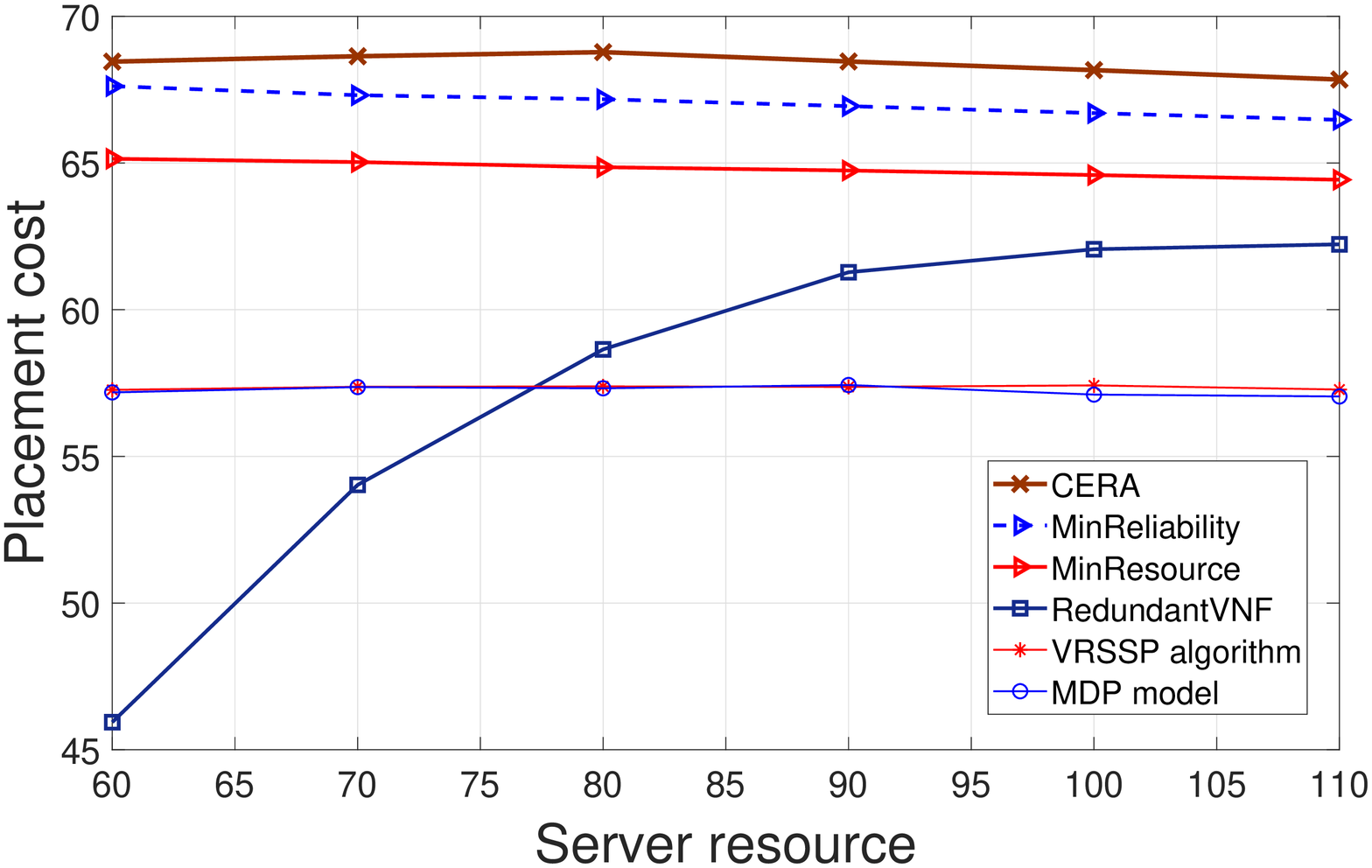}
		\vspace{-.1in}
		\renewcommand{\captionfont}{\small}
		\caption{\scriptsize{Placement cost for different values of server resource.}}
		\label{PerformanceEval_ServerResource_Cost}
	\end{subfigure}
	\begin{subfigure}[b]{0.495\textwidth}
		\includegraphics[scale=0.25]{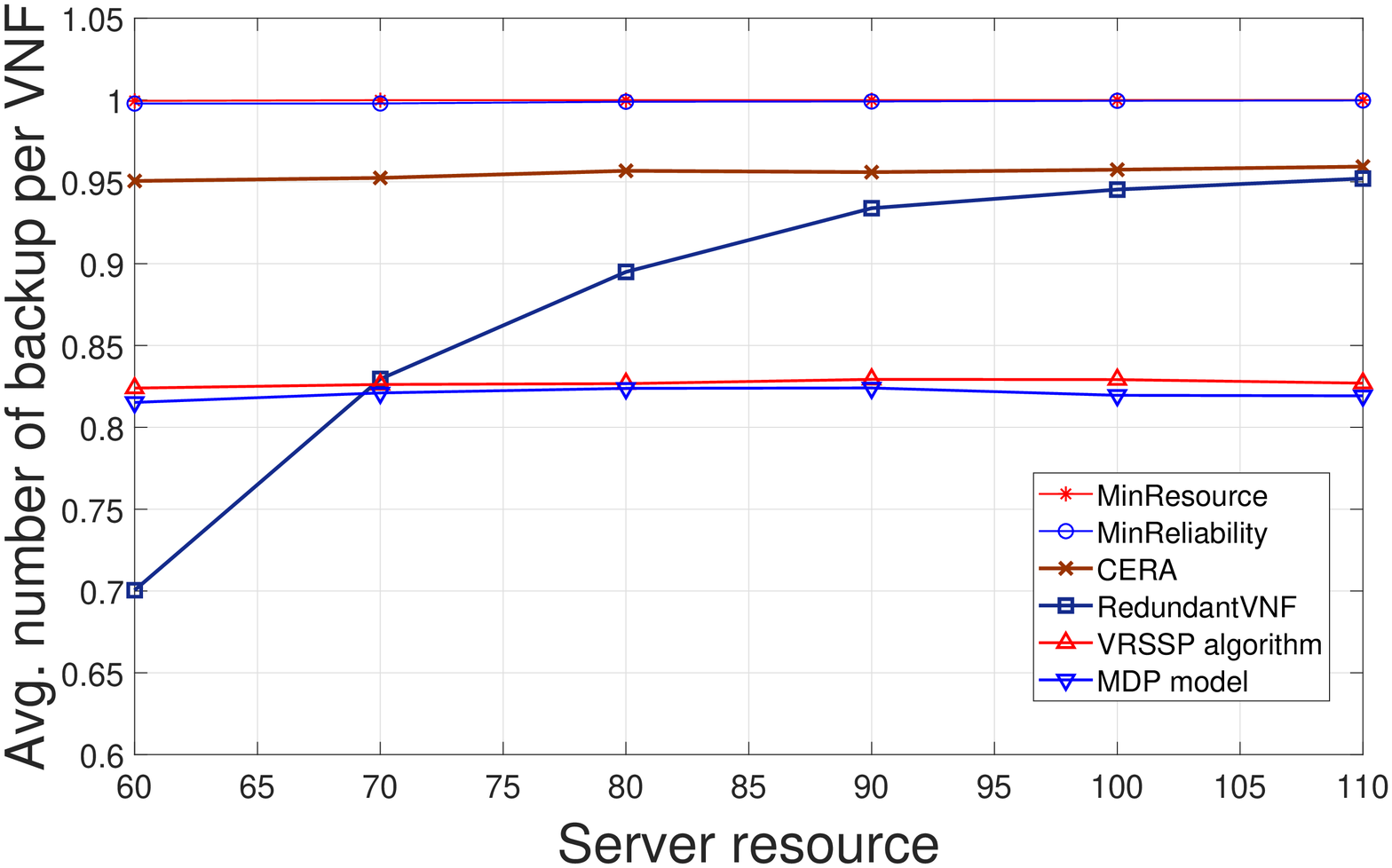}
		\vspace{-.1in}
		\renewcommand{\captionfont}{\small}
		\caption{\scriptsize{Number of backups per VNF for different values of server resourc.}}
		\label{PerformanceEval_ServerResource_BackupNum}
	\end{subfigure}
	\vspace{-5mm}
	\label{PerformanceEval_ServerResource}                            
	\caption{\small{Performance of MDP model for different values of server resource, $R^s_{i,j}$, compared to static methods.}}
	\vspace{-8mm}
\end{figure}

In Figs. \ref{PerformanceEval_ServerResource_AdmitRatio}-\ref{PerformanceEval_ServerResource_BackupNum}, the admission ratio, placement cost, and mean the number of backups per VNF for the MDP model and five static methods, by changing the values of server resource, $R^s_{i,j}$, are indicated, respectively.
The simulation is conducted with $d^l = 0.5, \beta = 15$. The value of $q^l$ is determined based on the number of the VNFs in the $l^{\text{th}}$ service type.
{\color{black}It is worth noting that we expect the advantage of the MDP is emerged in the full load scenario, in which the number of the incoming services is increased. Because when there exists sufficient resources in the InPs, most of the algorithms can admit the high percentage of the incoming services. For evaluating this fact, we fix the mean number of the incoming services and alter the amount of InPs' resources.}
{\color{black}According to Fig. \ref{PerformanceEval_ServerResource_AdmitRatio}, the admission ratio of the MDP model is significantly higher than the static models, especially for the low amount of server resources. This is due to the fact that by increasing the amount of server resources, most of the incoming services can be admitted by using a simple algorithm, and there is no need for more intelligence in service placement.
Also, as shown in Figs. \ref{PerformanceEval_ServerResource_Cost}-\ref{PerformanceEval_ServerResource_BackupNum}, the placement cost and the average number of backup servers in the MDP model are remarkably lower than the MinResource, MinReliability, CERA, and RedundantVNF methods and in the proximity of VRSSP algorithm. It is worth noting that the placement cost and the average number of backup servers of RedundantVNF are lower than the MDP model for the low amount of resources. Because the RedundantVNF method only admits the services with a low number of VNFs which leads to lower admission ratio compared to MDP model. However, RedundantVNF method still has a better performance in all metrics compared to MinResource, MinReliability, and CERA methods.}

\begin{figure}
	\centering
	\begin{subfigure}[b]{0.495\textwidth}
		\includegraphics[scale=0.24]{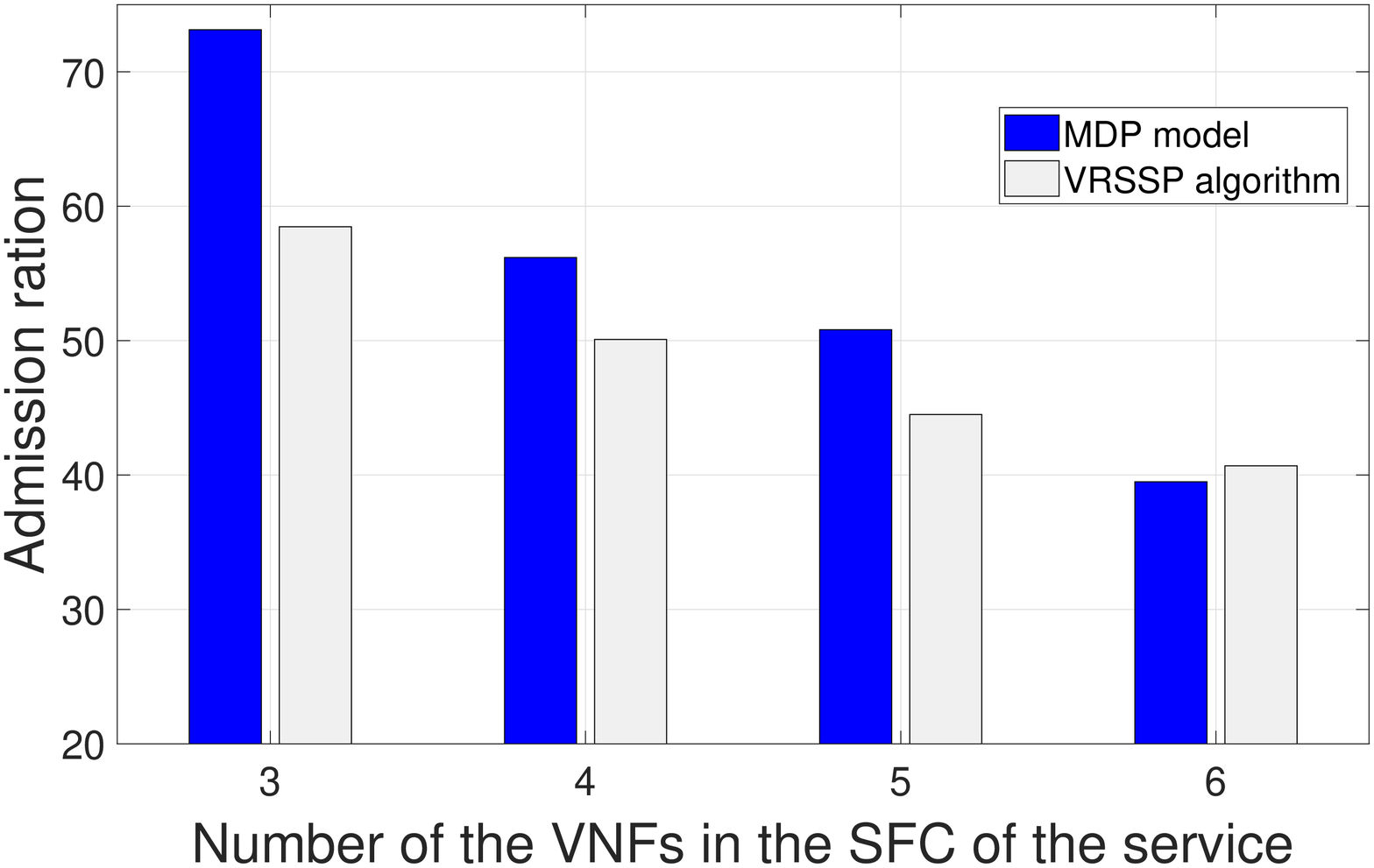}
		\vspace{-.1in}
		\renewcommand{\captionfont}{\small}
		\caption{\scriptsize{Admission ratio for different number of VNFs.}}
		\label{FairnessAnalysis_VNF_Number}
	\end{subfigure}
	\begin{subfigure}[b]{0.495\textwidth}
		\includegraphics[scale=0.24]{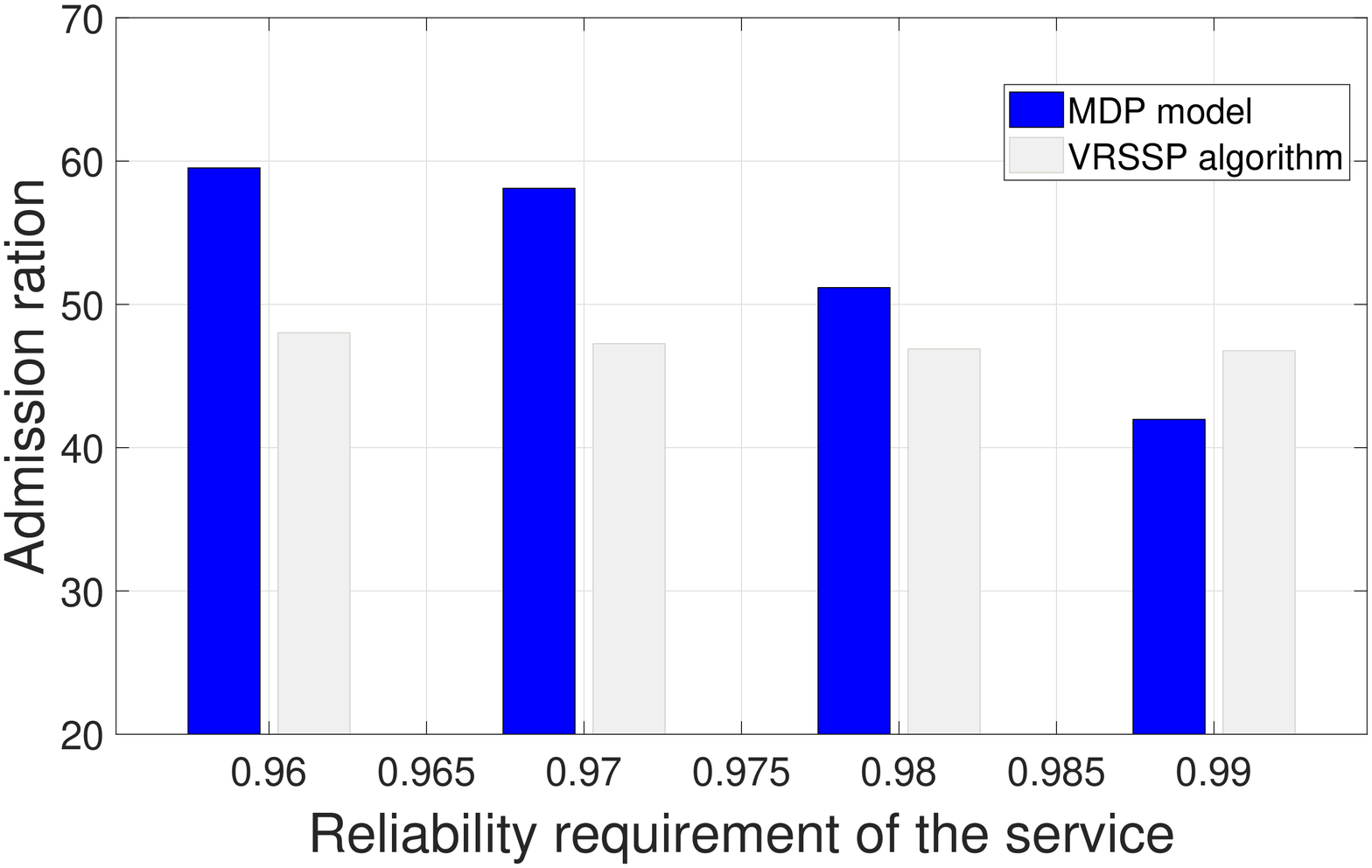}
		\vspace{-.1in}
		\renewcommand{\captionfont}{\small}
		\caption{\scriptsize{Placement cost for different values of reliability requirement.}}
		\label{FairnessAnalysis_ReliabilityRequirement}
	\end{subfigure}
	\vspace{-12mm}
	\label{FairnessAnalysis}                            
	\caption{\small{Admission ratio of the MDP model and VRSSP algorithm for different characteristics of the service types.}}
	\vspace{-10mm}
\end{figure}
{\color{black}One of the most important aspects of evaluating the performance of the MDP model is fairness analysis of this model. The most critical characteristics of the services which should be taken into consideration are the reliability requirement and number of VNFs of the services. For this purpose, in Figs. \ref{FairnessAnalysis_VNF_Number}-\ref{FairnessAnalysis_ReliabilityRequirement}, the admission ratio of the MDP model and the VRSSP algorithm for the different number of the VNFs and reliability requirements of the service is indicated. It is worth noting that between the introduced static methods, the VRSSP algorithm has the best performance. The simulation is conducted with $d^l = 0.5, \beta = 15, R^s_{i,j}=70$. The value of $q^l$ is determined based on the number of the VNFs in the $l^{\text{th}}$ service type.
As shown in Fig. \ref{FairnessAnalysis_VNF_Number}, the admission ratio of the MDP model is greater than the VRSSP algorithm when the number of VNF is $3, 4, 5$, and only when the number of VNF is $6$, the performance of VRSSP is slightly better than the MDP model. In other words, the MDP model can significantly improve the admission ratio in the cost of slightly reducing the admitted services with a high number of VNFs. There is similar observation by increasing the reliability requirement, as seen in Fig. \ref{FairnessAnalysis_ReliabilityRequirement}, indicates that the MDP model can significantly improve the admission ratio in the cost of slightly reducing the admitted services with high reliability requirement.
}

{\color{black}While our simulation results demonstrate the proof of concept, the accuracy of the simulations is one of the most important aspects of validating the paper. For this purpose, the repeatability of the simulations and the time horizon of terminating simulations are two key factors, as discussed in \cite{pawlikowski2002credibility, sarkar2014revisiting}.
For fulfilling the first aspect, we used Matlab software, version of 2017.b, which benefits the use of Mersenne Twister, one of the best pseudo-random number generators (PRNG) \cite{matsumoto1998mersenne, pawlikowski2002credibility}.
    Regarding the second aspect, we should discuss two points. First, for determined service types and NFVI, VVI algorithm runs to determine the optimal policy. Then, due to the random nature of service arrival and departure, a large-enough number of the slots should be simulated to evaluate the policy. The required number of slots is dependent on the size of state space, $|\Omega_S| = \prod_{l=1}^{L}{(\sigma^l_{\text{max}}+1)}\times\prod_{l=1}^{L}({\lambda^l_{\text{max}}+1)}$, which is equal to $6^4 \times 3^4 = 104976$, in the conducted simulation. Therefore, we simulated $10^6$ slots to have sufficient data for evaluating the policy. The second point is the existence of parameters which can change the output policy of VVI algorithm, for a given NFVI. In this way, the most effective parameters are the reliability requirement and the number of VNFs in each service type. In the considered simulation setup, there are $256$ different states for the reliability requirement, and also $256$ different states for the number of the VNFs of service types, which leads to $65536$ different combinations of these two parameters. However, a great number of these combinations is not meaningful. Nevertheless, simulating all of the combinations is time-consuming. Therefore, for each determined NFVI, we simulate about $1000$ different combinations. More precisely, for obtaining each point in Figs. 3-4 which has a determined NFVI, we simulated 1000 different combinations of reliability requirement and number of VNFs of the services. For each combination, we run VVI algorithm, then the resulted policy of VVI algorithm is evaluated during $10^6$ slots to determine the performance metrics. For a better clarification, we precisely report the admission ratio of MDP model and VRSSP algorithm which is the second best algorithm, when $R^s_{i,j} = 70$. The mean improvement of admission ratio using MDP model compared to VRSSP algorithm is $10.71\%$. The minimum and maximum improvement of admission ratio are $4.15\%$ and $18.74\%$.}

{\color{black}Finally, we compare the runtime of Algorithm \ref{Alg1} with MinResource, MinReliability, and CERA algorithms. All the simulations whose results are reported in this paper are conducted using a machine having an Intel 2.7 GHz processor and 8GB of RAM. Using this machine, the mean execution time of Algorithm \ref{Alg1} in each slot is 19 ms which is negligible compared to the length of the slot. The runtimes of MinResource, MinReliability, and CERA algorithms are 7 ms, 7 ms, and 12 ms, respectively. Because of solving an optimization problem in the RedundantVNF algorithm, the runtime of this method is significantly greater than the other methods.
}

\vspace{-4mm}
\section{Conclusion}
\label{Conclusion_Section}
In this paper, we investigated the reliability-aware service placement considering the dynamic nature of the service arrival and departure. We adopted a model based on an infinite horizon MDP for dynamic reliability-aware service placement, considering the simultaneous allocation of the main and backup servers. The reward function of the proposed MDP model was defined such that the admission ratio is maximized and placement cost is minimized.
For evaluating each action, we used a sub-optimal algorithm named VRSSP. Then, we introduced an algorithm named VVI, based on value iteration for finding the optimal policy. Finally, via extensive simulations, we compared the performance of the MDP model with five static methods
and demonstrated the superiority of the MDP model in terms of various criteria. Also, the robustness of the proposed MDP model in different scenarios and the fairness analysis is investigated.

\vspace{-6mm}

\bibliographystyle{ieeetran}
\bibliography{ref}

\begin{thebibliography}{10}
\providecommand{\url}[1]{#1}
\csname url@samestyle\endcsname
\providecommand{\newblock}{\relax}
\providecommand{\bibinfo}[2]{#2}
\providecommand{\BIBentrySTDinterwordspacing}{\spaceskip=0pt\relax}
\providecommand{\BIBentryALTinterwordstretchfactor}{4}
\providecommand{\BIBentryALTinterwordspacing}{\spaceskip=\fontdimen2\font plus
\BIBentryALTinterwordstretchfactor\fontdimen3\font minus
  \fontdimen4\font\relax}
\providecommand{\BIBforeignlanguage}[2]{{%
\expandafter\ifx\csname l@#1\endcsname\relax
\typeout{** WARNING: IEEEtran.bst: No hyphenation pattern has been}%
\typeout{** loaded for the language `#1'. Using the pattern for}%
\typeout{** the default language instead.}%
\else
\language=\csname l@#1\endcsname
\fi
#2}}
\providecommand{\BIBdecl}{\relax}
\BIBdecl

\bibitem{afolabi2018network}
I.~Afolabi, T.~Taleb, K.~Samdanis, A.~Ksentini, and H.~Flinck, ``Network
  slicing and softwarization: A survey on principles, enabling technologies,
  and solutions,'' \emph{IEEE Communications Surveys \& Tutorials}, vol.~20,
  no.~3, pp. 2429--2453, 2018.

\bibitem{mijumbi2016network}
R.~Mijumbi, J.~Serrat, J.-L. Gorricho, N.~Bouten, F.~De~Turck, and R.~Boutaba,
  ``Network function virtualization: State-of-the-art and research
  challenges,'' \emph{IEEE Communications Surveys \& Tutorials}, vol.~18,
  no.~1, pp. 236--262, 2016.

\bibitem{laghrissi2018survey}
A.~Laghrissi and T.~Taleb, ``A survey on the placement of virtual resources and
  virtual network functions,'' \emph{IEEE Communications Surveys \& Tutorials},
  vol.~21, no.~2, pp. 1409--1434, 2018.

\bibitem{herrera2016resource}
J.~G. Herrera and J.~F. Botero, ``Resource allocation in {NFV}: A comprehensive
  survey,'' \emph{IEEE Trans. on Network and Service Management}, vol.~13,
  no.~3, pp. 518--532, 2016.

\bibitem{osseiran2014scenarios}
A.~Osseiran, F.~Boccardi, V.~Braun, K.~Kusume, P.~Marsch, M.~Maternia,
  O.~Queseth, M.~Schellmann, H.~Schotten, H.~Taoka \emph{et~al.}, ``Scenarios
  for 5{G} mobile and wireless communications: the vision of the metis
  project,'' \emph{IEEE Communications Magazine}, vol.~52, no.~5, pp. 26--35,
  2014.

\bibitem{pham2017traffic}
C.~Pham, N.~H. Tran, S.~Ren, W.~Saad, and C.~S. Hong, ``Traffic-aware and
  energy-efficient {V}{N}{F} placement for service chaining: Joint sampling and
  matching approach,'' \emph{IEEE Trans. on Services Computing}, 2017.

\bibitem{bari2016orchestrating}
F.~Bari, S.~R. Chowdhury, R.~Ahmed, R.~Boutaba, and O.~C. M.~B. Duarte,
  ``Orchestrating virtualized network functions,'' \emph{IEEE Trans. on Network
  and Service Management}, vol.~13, no.~4, pp. 725--739, 2016.

\bibitem{mechtri2016scalable}
M.~Mechtri, C.~Ghribi, and D.~Zeghlache, ``A scalable algorithm for the
  placement of service function chains,'' \emph{IEEE Trans. on Network and
  Service Management}, vol.~13, no.~3, pp. 533--546, 2016.

\bibitem{herker2015data}
S.~Herker, X.~An, W.~Kiess, S.~Beker, and A.~Kirstaedter, ``Data-center
  architecture impacts on virtualized network functions service chain embedding
  with high availability requirements,'' in \emph{Workshop. of IEEE Globecom},
  San Diego, CA, Dec. 2015.

\bibitem{fan2015grep}
J.~Fan, Z.~Ye, C.~Guan, X.~Gao, K.~Ren, and C.~Qiao, ``Grep: Guaranteeing
  reliability with enhanced protection in {NFV},'' in \emph{Proc. of ACM
  SIGCOMM Workshop}, Heraklion, Greece, July. 2015.

\bibitem{ding2017enhancing}
W.~Ding, H.~Yu, and S.~Luo, ``Enhancing the reliability of services in {NFV}
  with the cost-efficient redundancy scheme,'' in \emph{Proc. of IEEE ICC},
  Paris, France, May. 2017.

\bibitem{fan2018framework}
J.~Fan, M.~Jiang, O.~Rottenstreich, Y.~Zhao, T.~Guan, R.~Ramesh, S.~Das, and
  C.~Qiao, ``A framework for provisioning availability of {NFV} in data center
  networks,'' \emph{IEEE JSAC}, vol.~36, no.~10, pp. 2246--2259, 2018.

\bibitem{qu2017reliability}
L.~Qu, C.~Assi, K.~Shaban, and M.~J. Khabbaz, ``A reliability-aware network
  service chain provisioning with delay guarantees in {NFV}-enabled enterprise
  datacenter networks,'' \emph{IEEE TNSM}, vol.~14, no.~3, pp. 554--568, 2017.

\bibitem{sun2018reliability}
J.~Sun, G.~Zhu, G.~Sun, D.~Liao, Y.~Li, A.~K. Sangaiah, M.~Ramachandran, and
  V.~Chang, ``A reliability-aware approach for resource efficient virtual
  network function deployment,'' \emph{IEEE Access}, vol.~6, pp.
  18\,238--18\,250, 2018.

\bibitem{kanizo2017optimizing}
Y.~Kanizo, O.~Rottenstreich, I.~Segall, and J.~Yallouz, ``Optimizing virtual
  backup allocation for middleboxes,'' \emph{IEEE/ACM Trans. on Networking},
  vol.~25, no.~5, pp. 2759--2772, 2017.

\bibitem{kanizo2018designing}
------, ``Designing optimal middlebox recovery schemes with performance
  guarantees,'' \emph{IEEE Journal on Selected Areas in Communications},
  vol.~36, no.~10, pp. 2373--2383, 2018.

\bibitem{rottenstreich2016minimizing}
O.~Rottenstreich, I.~Keslassy, Y.~Revah, and A.~Kadosh, ``Minimizing delay in
  network function virtualization with shared pipelines,'' \emph{IEEE
  Transactions on Parallel and Distributed Systems}, vol.~28, no.~1, pp.
  156--169, 2016.

\bibitem{chen2019automated}
X.~Chen, W.~Ni, I.~B. Collings, X.~Wang, and S.~Xu, ``Automated function
  placement and online optimization of network functions virtualization,''
  \emph{IEEE Trans. on Communications}, vol.~67, no.~2, pp. 1225--1237, 2019.

\bibitem{chen2018multi}
X.~Chen, W.~Ni, T.~Chen, I.~Collings, X.~Wang, R.~P. Liu, and G.~B. Giannakis,
  ``Multi-timescale online optimization of network function virtualization for
  service chaining,'' \emph{IEEE Trans. on Mobile Computing}, 2018.

\bibitem{woldeyohannes2018cluspr}
Y.~T. Woldeyohannes, A.~Mohammadkhan, K.~Ramakrishnan, and Y.~Jiang, ``Cluspr:
  Balancing multiple objectives at scale for {NFV} resource allocation,''
  \emph{IEEE Trans. on Network and Service Management}, vol.~15, no.~4, pp.
  1307--1321, 2018.

\bibitem{yang2018cost}
B.~Yang, W.~K. Chai, Z.~Xu, K.~V. Katsaros, and G.~Pavlou, ``Cost-efficient
  {NFV}-enabled mobile edge-cloud for low latency mobile applications,''
  \emph{IEEE Trans. on Network and Service Management}, vol.~15, no.~1, pp.
  475--488, 2018.

\bibitem{cao2019dynamic}
H.~Cao, H.~Zhu, and L.~Yang, ``Dynamic embedding and scheduling of service
  function chains for future {SDN}/{NFV}-enabled networks,'' \emph{Accepted in
  IEEE Access}, 2019.

\bibitem{jia2018online}
Y.~Jia, C.~Wu, Z.~Li, F.~Le, A.~Liu, Z.~Li, Y.~Jia, C.~Wu, F.~Le, and A.~Liu,
  ``Online scaling of {NFV} service chains across geo-distributed
  datacenters,'' \emph{IEEE/ACM Trans. on Networking (TON)}, vol.~26, no.~2,
  pp. 699--710, 2018.

\bibitem{eramo2017approach}
V.~Eramo, E.~Miucci, M.~Ammar, and F.~G. Lavacca, ``An approach for service
  function chain routing and virtual function network instance migration in
  network function virtualization architectures,'' \emph{IEEE/ACM Transactions
  on Networking}, vol.~25, no.~4, pp. 2008--2025, 2017.

\bibitem{eramo2017migration}
V.~Eramo, M.~Ammar, and F.~G. Lavacca, ``Migration energy aware
  reconfigurations of virtual network function instances in {NFV}
  architectures,'' \emph{IEEE Access}, vol.~5, pp. 4927--4938, 2017.

\bibitem{liu2017dynamic}
J.~Liu, W.~Lu, F.~Zhou, P.~Lu, and Z.~Zhu, ``On dynamic service function chain
  deployment and readjustment,'' \emph{IEEE Trans. on Network and Service
  Management}, vol.~14, no.~3, pp. 543--553, 2017.

\bibitem{khezri2018deep}
H.~R. Khezri, P.~A. Moghadam, M.~K. Farshbafan, V.~Shah-Mansouri, H.~Kebriaei,
  and D.~Niyato, ``Deep {Q}-{L}earning for dynamic reliability aware
  {NFV}-based service provisioning,'' \emph{arXiv preprint arXiv:1812.00737},
  2018.

\bibitem{shafiq2011characterizing}
M.~Z. Shafiq, L.~Ji, A.~X. Liu, and J.~Wang, ``Characterizing and modeling
  internet traffic dynamics of cellular devices,'' \emph{ACM SIGMETRICS
  Performance Evaluation Review}, vol.~39, no.~1, pp. 265--276, 2011.

\bibitem{kaelbling1998planning}
L.~P. Kaelbling, M.~L. Littman, and A.~R. Cassandra, ``Planning and acting in
  partially observable stochastic domains,'' \emph{Artificial intelligence},
  vol. 101, no. 1-2, pp. 99--134, 1998.

\bibitem{google_SLA}
``Google apps service level agreement,'' [Online].
  Available:http://www.google.com/apps/intl/en/terms/sla.html.

\bibitem{pawlikowski2002credibility}
K.~Pawlikowski, H.-D. Jeong, and J.-S. Lee, ``On credibility of simulation
  studies of telecommunication networks,'' \emph{IEEE Communications magazine},
  vol.~40, no.~1, pp. 132--139, 2002.

\bibitem{sarkar2014revisiting}
N.~I. Sarkar and J.~A. Guti{\'e}rrez, ``Revisiting the issue of the credibility
  of simulation studies in telecommunication networks: highlighting the results
  of a comprehensive survey of ieee publications,'' \emph{IEEE Communications
  Magazine}, vol.~52, no.~5, pp. 218--224, 2014.

\bibitem{matsumoto1998mersenne}
M.~Matsumoto and T.~Nishimura, ``Mersenne twister: a 623-dimensionally
  equidistributed uniform pseudo-random number generator,'' \emph{ACM
  Transactions on Modeling and Computer Simulation (TOMACS)}, vol.~8, no.~1,
  pp. 3--30, 1998.

\end{thebibliography}
\vspace{-6mm}
\appendix[Procedure of the VRSSP algorithm]
\label{appendix_VRSSP}
The two important aspects of the Viterbi-based algorithm are definition of state and decision metric. In the service placement problem, the state of each stage is defined as a set of the servers where respective VNF can run. Therefore, the set of states in the $m^{\text{th}}$ stage, $X_m$, is
\vspace{-2mm}
\begin{align}
X_m=\begin{cases} \{0\} \bigcup G^s, & (m \bmod 2) = 0, \\ G^s, & (m \bmod 2) = 1,\end{cases} \quad m =1,\ldots,L_V,
\label{State_Determination}
\end{align}
where $G^s$ is the set of all servers introduced in Section \ref{InP_SubSec} and $0$ indicates no server assignment which is applicable only in even stages. For defining decision metric, we exploit the placement cost and cost of violating reliability requirement. Let $\Theta_{m-1,m}^{x_1,x_2}$ denote the transition cost between the $x_1^{\text{th}}$ state of $(m-1)^{\text{th}}$ stage and $x_2^{\text{th}}$ state of $m^{\text{th}}$ stage which can be computed as
\vspace{-3mm}
\begin{align}
\Theta_{m-1,m}^{x_1,x_2}&= \textstyle \sum_{j=1}^{|R|} r^{\rho_{k_m}}_{u_m,j}\times C_{i_m^{x_2},j} + DC_{i_m^{x_2},t_{u_m}^{\rho_{k_m}}} + \psi_{m-1,m}^{x_1,x_2,r} + \psi_{m-1,m}^{x_1,x_2,b} + \phi_{m-1}^{x_1},
\label{DecisionMetric}
\end{align}
where $r^{\rho_{k_m}}_{u_m,j}$ is the amount of $j^{\text{th}}$ resource type required for the $u_m^{\text{th}}$ VNF of $k_m^{\text{th}}$ service, $\rho_{k_m}$ refers to the type of $k_m^{\text{th}}$ service and $C_{i_m^{x_2},j}$ is the unit cost of using $j^\text{th}$ resource type of $(i_m^{x_2})^\text{th}$ InP's servers. Also, $k_m$ denote the index of considered service in the $m^{\text{th}}$ stage, $u_m$ is the index of considered VNF of $k_m^{\text{th}}$ service in the $m^{\text{th}}$ stage and $i_m^{x_2}$ indicates the InP index for the $x_2^{\text{th}}$ state of $m^{\text{th}}$ stage.  $DC_{i_m^{x_2},t_{u_m}^{\rho_{k_m}}}$ refers to the deployment cost of $(t_{u_m}^{\rho_{k_m}})^{\text{th}}$ types of VNF in the servers of $(i_m^{x_2})^{\text{th}}$ InP. Also, $\phi_{m-1}^{x_1}$ indicates the cost of being in the $x_1^{\text{th}}$ state of $(m-1)^{\text{th}}$ stage. The $\psi_{m-1,m}^{x_1,x_2, b}$ is the cost of traffic routing for the transition between the $x_1^{\text{th}}$ state of $(m-1)^{\text{th}}$ stage and the $x_2^{\text{th}}$ state of $m^{\text{th}}$ stage which can be computed as
\vspace{-2mm}
\begin{align}
&\psi_{m-1,m}^{x_1,x_2, b}=
\begin{cases}0 & u_m=1 \; \text{or} \; x_2=0, \\
b^{\rho_{k_m}} \times \Big(C_{i_{m-1}^{x_1,2},i_m^{x_2}}^{s_{m-1}^{x_1,2},s_m^{x_2}} + C_{i_{m-1}^{x_1,1},i_m^{x_2}}^{s_{m-1}^{x_1,1},s_m^{x_2}}\Big) & u_m\geq2, m \in \text{odds},  \\
b^{\rho_{k_m}} \times \Big(C_{i_{m-1}^{x_1,2},i_m^{x_2}}^{s_{m-1}^{x_1,2},s_m^{x_2}} + C_{i_{m-1}^{x_1,3},i_m^{x_2}}^{s_{m-1}^{x_1,3},s_m^{x_2}}\Big) & u_m\geq2, m \in \text{evens}, \end{cases} \label{LC_Calculation}
\end{align}
where we have $i_{m-1}^{x_1,1} = \Lambda_{m-1,x_1}^{I}[m-1]$, $s_{m-1}^{x_1,1} = \Lambda_{m-1,x_1}^{W}[m-1]$, $i_{m-1}^{x_1,2} = \Lambda_{m-1,x_1}^{I}[m-2]$, $s_{m-1}^{x_1,2} = \Lambda_{m-1,x_1}^{W}[m-2]$, $i_{m-1}^{x_1,3} = \Lambda_{m-1,x_1}^{I}[m-3]$, and $s_{m-1}^{x_1,3} = \Lambda_{m-1,x_1}^{W}[m-3]$.
The parameters $\Lambda_{m-1,x_1}^I$ and $\Lambda_{m-1,x_1}^W$ are the vectors of length $m-1$ which indicate the index of InP and server for the survived path of $x_1^{\text{th}}$ state in the $(m-1)^{\text{th}}$ stage. Also, $s_m^{x_2}$ indicates the index of server in the $x_2^{\text{th}}$ state of $m^{\text{th}}$ stage. When $x_2=0$, we assume $i_m^{x_2}=s_m^{x_2}=0$ and consider $C_{0,j}=0, \; C_{i_{m-1}^{x_1,1}, 0}^{s_{m-1}^{x_1,1}, 0} = 0, \; C_{i_{m-1}^{x_1,2}, 0}^{s_{m-1}^{x_1,2}, 0} = 0, \; C_{i_{m-1}^{x_1,3}, 0}^{s_{m-1}^{x_1,3}, 0} = 0$. Also, we have $C^{0,s_m^{x_2}}_{0,i_m^{x_2}} = 0$.

Finally, $\psi_{m-1,m}^{x_1,x_2, r}$ is the cost of violating reliability requirement for the transition between the $x_1^{\text{th}}$ state of $(m-1)^{\text{th}}$ stage and the $x_2^{\text{th}}$ state of $m^{\text{th}}$ stage which can be computed as
\vspace{-3mm}
\begin{align}
\psi_{m-1,m}^{x_1,x_2, r} =M^{\rho_{k_m}} \times\big(OR_m^{x_2}-T_{m-1,m}^{x_1,x_2}\big)\times I\big(OR_m^{x_2}-T_{m-1,m}^{x_1,x_2}\big),
\end{align}
where, $M^{\rho_{k_m}}$ is the cost of violating the reliability requirement of $(\rho_{k_m})^{\text{th}}$ service type. It is worth noting that the value of $M^{\rho_{k_m}}$ should be selected large enough compared to the placement cost to meet the reliability requirement. The parameter $OR_m^{x_2}$ is the objective reliability in the $x_2^{\text{th}}$ state of $m^{\text{th}}$ stage which is defined as $OR_m^{x_2} = 1$ when $x_2 = 0$ and otherwise $OR_m^{x_2} = 1-F^{\rho_{k_m}}$. In $x_2=0$ which means no backup server assignment for the considered VNF, the path which leads to the maximum reliability is selected as a survived path. $T_{m-1,m}^{x_1,x_2}$ is defined as the reliability level in the $x_2^{\text{th}}$ state of $m^{\text{th}}$ stage, if the VRSSP algorithm moves from the $x_1^{\text{th}}$ state of $(m-1)^{\text{th}}$ stage to the $x_2^{\text{th}}$ state of $m^{\text{th}}$ stage. The value of $T_{m-1,m}^{x_1,x_2}$, can be computed as
\vspace{-4mm}
\begin{align}
&T_{m-1,m}^{x_1,x_2}= \begin{cases} 1-v_{i_m^{x_2}} & u_m=1, m \in \text{odds}, \\ 1-\big( v_{\Lambda_{m-1,x_1}^{I}[m-1]}\times v_{i_m^{x_2}} \big) & u_m=1, m \in \text{evens}, \\ \tau_{m-1}^{x_1} \times (1-v_{i_m^{x_2}}) & u_m\geq2, m \in \text{odds},
\\ \dfrac{\tau_{m-1}^{x_1} \times \big(1-(v_{\Lambda_{m-1,x_1}^{I}[m-1]} \times v_{i_m^{x_2}})\big)}{1-v_{\Lambda_{m-1,x_1}^{I}[m-1]}} & u_m\geq2, m \in \text{evens}, \end{cases} \label{TR_Calculation}
\end{align}
where $\tau^{x_1}_{m-1}$ is the reliability of being in the $x_1^{\text{th}}$ state of $(m-1)^{\text{th}}$ stage.

The survived path for the $x_2^{\text{th}}$ state of $m^{\text{th}}$ stage can be computed using the decision metric as
\vspace{-12mm}
\begin{align}
\textstyle {SP}_{m}^{x_2} &= \underset{x_1 \, \in \, XP_{m}^{x_2}}{\operatorname{argmin}} \big(\Theta_{m-1,m}^{x_1,x_2}\big), \label{SP_Selection}\\
\textstyle \Lambda_{m,x_2}^I &= \left[\begin{array}{cc} \Lambda_{m-1,{SP}_{m}^{x_2}}^{I} & i_m^{x_2} \end{array}\right], \;
\textstyle \Lambda_{m,x_2}^W = \left[\begin{array}{cc} \Lambda_{m-1,{SP}_{m}^{x_2}}^{W} & s_m^{x_2} \end{array}\right] \label{InP_Server_index},
\end{align}
where ${SP}_{m}^{x_2}$ indicates the index of the survived path and $XP_{m}^{x_2}$ is the set of possible input states to $x_2^{\text{th}}$ state of $m^{\text{th}}$ stage. The $XP_{m}^{x_2}$ is the subset of the states set in the $(m-1)^{\text{th}}$ stage $(XP_{m}^{x_2} \subseteq X_{m-1})$ in which there are enough resources for running the VNF of $m^{\text{th}}$ stage in the server of $x_2^{\text{th}}$ state. In \eqref{InP_Server_index}, the characteristic of survived path in the $x_2^{\text{th}}$ state of $m^{\text{th}}$ stage is determined in which $i_m^{x_2}$ and $s_m^{x_2}$ indicate the index of the considered InP and server in the $x_2^{\text{th}}$ state of $m^{\text{th}}$ stage, respectively. The values of $\phi_m^{x_2}$ and $\tau_m^{x_2}$ can be, respectively, determined as 
\vspace{-4mm}
\begin{align}
\phi_{m}^{x_2} &= \phi_{m-1}^{SP_{m}^{x_2}} + \Theta_{m-1,m}^{SP_{m}^{x_2},x_2} -\psi_{m-1,m}^{SP_{m}^{x_2},x_2, r} , \label{SC_Update} \\
\tau_{m}^{x_2} &= \begin{cases} 1-v_{i_m^{x_2}} & u_m=1, m \in \text{odds}, \\
1-\big( v_{\Lambda_{m,x_2}^{I}[m-1]}\times v_{i_m^{x_2}} \big) & u_m=1, m \in \text{evens}, \\
\tau_{m-1}^{SP_m^{x_2}} \times (1-v_{i_m^{x_2}}) & u_m\geq2, m \in \text{odds}, \\
\dfrac{\tau_{m-1}^{SP_m^{x_2}}  \big(1-(v_{\Lambda_{m,x_2}^{I}[m-1]} \times v_{i_m^{x_2}})\big)}{1-v_{\Lambda_{m,x_2}^{I}[m-1]}} & u_m\geq2, m \in \text{evens}. \end{cases} \label{SR_Update}
\end{align}

In the last stage, $L_V$, the best path between the survived paths can be determined using
\vspace{-3mm}
\begin{align}
\textstyle \zeta &=  \underset{x \, \in \, X_{L_V}}{\operatorname{argmin}} \Big(\phi_{L_V}^{x} + M^{\rho_{K}}\times\big(OR_{L_V}^{x}-\tau_{L_V}^{x}\big)\times I\big(OR_{L_V}^{x}-\tau_{L_V}^{x}\big)\Big), \notag \\ \textstyle \mathcal{P}^I_{L_V} &= \Lambda_{L_V,\zeta}^{I}, \; \mathcal{P}^W_{L_V} = \Lambda_{L_V,\zeta}^{W}, \label{BP_Calculation}
\end{align}
where $\mathcal{P}^I_{L_V}$ and $\mathcal{P}^W_{L_V}$ are the indexes of the InP and server with the best path, respectively, and $\zeta$ is the index of best path in the last stage. The number of candidates for the best path is $\big|X_{L_V}\big|$.\\

%
%
%

\end{document}